# Fresnel reflection and transmission in the presence of gain media

Masud Mansuripur[†] and Per K. Jakobsen[‡]

[†]James C. Wyant College of Optical Sciences, The University of Arizona, Tucson, Arizona, USA
[‡]Department of Mathematics and Statistics, UIT The Arctic University of Norway, Tromsø, Norway

**Abstract**. When a monochromatic electromagnetic plane-wave arrives at the flat interface between its transparent host (i.e., the incidence medium) and an amplifying (or gainy) second medium, the incident beam splits into a reflected wave and a transmitted wave. In general, there is a sign ambiguity in connection with the $k$-vector of the transmitted beam, which requires at the outset that one decide whether the transmitted beam should grow or decay as it recedes from the interface. The question has been posed and addressed most prominently in the context of incidence at large angles from a dielectric medium of high refractive index onto a gain medium of lower refractive index. Here, the relevant sign of the transmitted $k$-vector determines whether the evanescent-like waves within the gain medium exponentially grow or decay away from the interface. We examine this and related problems in a more general setting, where the incident beam is taken to be a finite-duration wavepacket whose footprint in the interfacial plane has a finite width. Cases of reflection from and transmission through a gainy slab of finite-thickness as well as those associated with a semi-infinite gain medium will be considered. The broadness of the spatio-temporal spectrum of our incident wavepacket demands that we develop a general strategy for deciding the signs of all the $k$-vectors that enter the gain medium. Such a strategy emerges from a consideration of the causality constraint that is naturally imposed on both the reflected and transmitted wavepackets.

**1. Introduction**. In a recent paper,[1] we described a theoretical approach to solving the problem of optical reflection and transmission at the interface between an ordinary dielectric medium and a linear, homogeneous, isotropic, weakly-amplifying second medium. The case in which the second (gain) medium is semi-infinite as well as that in which the gainy material is a slab of finite thickness sandwiched between two semi-infinite ordinary dielectric media were discussed in detail; see Fig.1. We also reported the results of extensive numerical calculations that demonstrated the viability of our theoretical approach to studying Fresnel reflection and transmission in the presence of gain media. One requirement for the analysis of optical reflection/transmission in the presence of gain is to restrict the incident beam to one that has a finite footprint at the interface with the (passive) incidence medium, lest the amplified waves arriving from far away regions obscure or contaminate the waves generated locally (or contributed by nearby regions of the gain medium). Similarly, it is essential to ensure that the incident beam has a well-defined time of arrival at the interface, say, $t = t_0$, so that no light, no matter how weak, could possibly have reached the gain medium prior to this time and been amplified to such an extent that it now contaminates the waves generated by the light that arrives after $t = t_0$.

The twin constraints of a finite footprint and a well-defined starting time for the incident beam necessitate a spectral decomposition of the incident wavepacket into its plane-wave constituents that are subsequently identified by their respective propagation vector $\boldsymbol{k}$ and temporal frequency $\omega$. Traditionally, such spectral decompositions are carried out using standard Fourier transformation. Denoting by $xy$ the interfacial plane at the junction of the incidence and transmittance media, the Fourier transform of the incident field at the $xy$-plane of the interface will be a function of the real-valued parameters $k_x$, $k_y$, and $\omega$. Whereas such spectral decompositions are totally adequate for the analysis of systems comprised exclusively of passive (i.e., transparent or partially absorptive) media, it turns out that the presence of optical gain in the system demands a generalization of the theoretical methods involving an extension of $k_x$, $k_y$, and/or $\omega$ into the complex plane. The resulting transformations are broadly classified as the Fourier-Laplace transforms. In [1], we used Fourier-Laplace transformations along with methods of complex-function analysis to investigate the reflection and transmission of light in the presence of gain media in the case of a single surface (or a single interface) as well as that of a finite-thickness gainy slab. The goal of the present paper is to explain, or to further elucidate, some of the underlying mathematical arguments that did not make it into [1], in part due to the limited available space, but also because some of these arguments were deemed disruptive to the logical flow of reasoning that was needed to bring about an understanding of the numerical results that we reported in that paper.

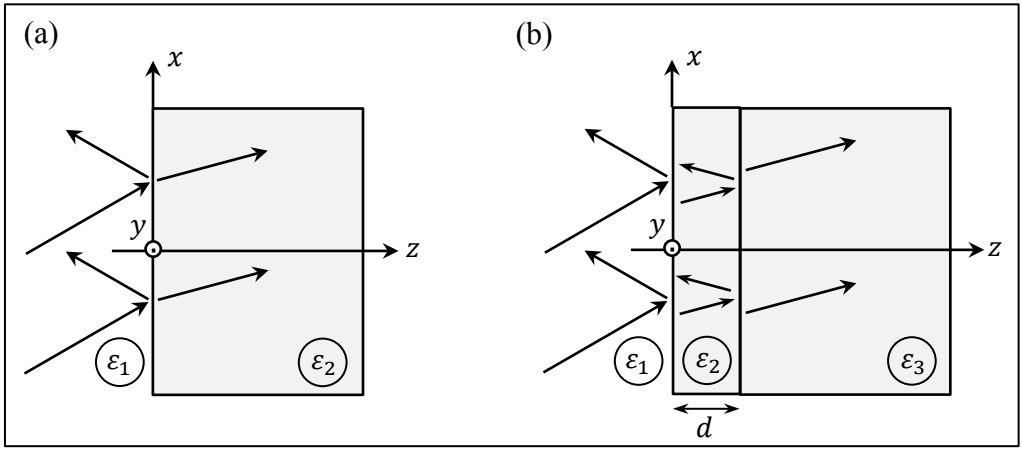

**Fig.1**. Reflection and transmission of light at the interface between a passive host medium (medium 1) and a gain medium. (a) The gain medium (medium 2) is semi-infinite. (b) The gain medium 2 is a slab of finite thickness $d$, sandwiched between medium 1 and a second semi-infinite passive medium 3. All three media are linear, homogeneous, isotropic, and non-magnetic, having relative permeability $\mu(\omega) = 1$; their optical properties are fully specified by their relative permittivities $\varepsilon(\omega)$. The incident beam (or light pulse) is a compact wavepacket of finite footprint along the $x$-axis and finite duration. To simplify the analysis, it is assumed that the incident beam as well as media 1, 2, and 3 are infinitely wide and uniform along the $y$-axis.

The present paper is organized as follows. In Sec. 2, we describe the application of the Fourier-Laplace transformation to a simple problem of electromagnetic wave propagation in free space. Here, we see how an argument based on the principle of causality[2,3] can be used to decide the sign of the $z$-component $k_z$ of the propagation vector $\boldsymbol{k}$, once the temporal frequency $\omega$ and the components $k_x$ and $k_y$ of a plane-wave's $k$-vector are specified. To apply the powerful principle of causality, we need to invoke the analyticity of the various functions involved, which requires a careful examination of the branch-cuts and branch-points of certain complex functions.[4-6] Appendix A provides a brief description of the notion of branch-cuts in complex analysis, while Appendix B examines the trajectories of the branch-points of $k_z$ in the complex $\omega$-plane. The reader who is not familiar with the Fourier-Laplace transformation may consult Appendix C for an overview of the basic mathematics as well as some of the elementary properties of this transformation.

Section 3 provides a detailed discussion of the poles, zeros, branch-points, and branch-cuts of $k_z$ for a plane-wave propagating inside a linear, homogeneous, isotropic, and non-magnetic medium whose optical properties are specified by a single-Lorentz-oscillator model of its dielectric permittivity $\varepsilon(\omega)$.[2,3,7,8] Both passive (i.e., transparent or lossy) and active (i.e., gainy) material media will be covered in this section, and the fundamental difference between passive and active media—pertaining to the domain of analyticity of $k_z$ and the role played by the branch-point trajectories in establishing such domains of analyticity — will be explored. The all-important and universal causality constraint,[2,3] which ultimately dictates the correct sign for each $k_z$ through a judicious choice of the branch-cuts, is the subject of Sec. 4.

In Sec. 5, we present an elementary model for a finite-footprint and finite-duration wavepacket that serves as a starting point for a discussion of the (spatio-temporal) spectral decomposition of the incident beam. Spectral decomposition enables us to treat the problem of reflection/transmission at a flat interface as a simpler problem involving a superposition of plane-waves, with each such plane-wave being specified in terms of a propagation vector $\boldsymbol{k}$ and a temporal frequency $\omega$.[2,3] The finite duration and the finite footprint of the incident beam inevitably result in an infinitely wide spectral profile, the reason behind which is briefly described in Appendix D. Certain subtleties associated with the infinite bandwidth of the spectral distribution in the $(\boldsymbol{k}, \omega)$ space are pointed out in Appendix E.

In Sec. 6, we apply the Fourier-Laplace transformation to examine the reflected and transmitted wavepackets for the single-surface problem of Fig. 1(a). To simplify the analysis, the incident packet will be assumed to be infinitely-wide and uniform in the $y$-direction, and also linearly-polarized, having its $E$-field aligned with the $y$-axis — in other words, the incident packet is taken to be $s$-polarized.[7,8] It will be seen that the presence of gain in the (semi-infinite) medium 2 necessitates a deformation of one of the integration contours: Either the inverse Fourier-Laplace integral must be evaluated along a displaced contour in the complex $\omega$-plane, in which case the integration path in the $k_x$-plane remains along the real axis; or the inverse transform integral in the $\omega$-plane is computed along the real axis, in which case the $k_x$-plane integration path must be deformed. These two methods are equivalent, and all numerical results based on them should turn out to be identical.

The problem of reflection/transmission from the finite-thickness slab of Fig. 1(b) is taken up in Sec. 7. Once again, the Fourier-Laplace transformation method will be called upon to deal with the necessity of deforming one of the integration contours. Unlike the single-surface problem of Sec. 6, here the Fresnel reflection and transmission coefficients of the slab turn out to have singularities (i.e., poles) in the upper-half $\omega$-plane. We explain how a proper deformation of the integration path away from the real axis of the complex $k_x$-plane removes the aforementioned singularities, thus paving the way for a return to the real axis of the $\omega$-plane without violating the causality constraint. The paper closes with a summary of the results and a few concluding remarks in Sec. 8.

**2. An application example of the Fourier-Laplace transformation**. With reference to Fig. 2(a), let the electromagnetic beam be initially defined in the $xy$-plane at $z = 0$, where the beam's $E$-field amplitude distribution is given by

$$\boldsymbol{E}(x, y, z = 0, t) = E_x(x, y, t)\hat{\boldsymbol{x}} + E_y(x, y, t)\hat{\boldsymbol{y}} + E_z(x, y, t)\hat{\boldsymbol{z}}. \tag{1}$$

We shall be interested in beams that are turned on at $t = 0$, and would like to express them as superpositions of plane-waves $\boldsymbol{E}_0 \exp[\mathrm{i}(\boldsymbol{k} \cdot \boldsymbol{r} - \omega t)]$, where, in general, $\boldsymbol{E}_0 = \boldsymbol{E}'_0 + \mathrm{i}\boldsymbol{E}''_0$ is a complex-valued constant vector, $\boldsymbol{k} = k_x\hat{\boldsymbol{x}} + k_y\hat{\boldsymbol{y}} + k_z\hat{\boldsymbol{z}}$ is the propagation vector whose $k_x$ and $k_y$ components are real, but whose $k_z$ could be complex-valued, $\boldsymbol{r} = x\hat{\boldsymbol{x}} + y\hat{\boldsymbol{y}} + z\hat{\boldsymbol{z}}$ is the position vector in 3-dimensional (3D) Cartesian space, and $\omega = \omega' + \mathrm{i}\omega''$ is the complex frequency, with $\omega'' \geq 0$ allowing for the exponential growth of the beam amplitude as a function of time $t$.

In order to decompose the initial amplitude distribution into a superposition of such plane-waves, we perform a Fourier-Laplace transform on individual $E$-field components, as follows:

$$\mathcal{E}_x(k_x, k_y, \omega) = \iint_{x,y=-\infty}^{\infty} \int_{t=0}^{\infty} E_x(x, y, t) \exp[-\mathrm{i}(k_x x + k_y y - \omega t)] \,\mathrm{d}x\mathrm{d}y\mathrm{d}t. \tag{2}$$

Similar expressions may be written for $\mathcal{E}_y$ and $\mathcal{E}_z$ as well. In writing Eq.(2), we have assumed that $E_x(x, y, t)$ is sufficiently well-behaved as a function of the spatial coordinates $(x, y)$ that its 2D Fourier transform exists in the real-valued $(k_x, k_y)$ domain. Moreover, the temporal growth of $E_x(x, y, t)$ is assumed to be bounded by $\exp(\Omega t)$, where $\Omega \geq 0$ is some real-valued constant. Consequently, the Laplace transform in Eq.(2) exists provided that $\text{Imag}(\omega) = \omega'' \geq \Omega$. An inverse Fourier-Laplace transformation now yields

$$E_x(x, y, t) = \iint_{k_x, k_y = -\infty}^{\infty} \int_{\omega = -\infty + i\Omega}^{\infty + i\Omega} \mathcal{E}_x(k_x, k_y, \omega) \exp[i(k_x x + k_y y - \omega t)] \, dk_x dk_y d\omega. \tag{3}$$

The domain of integration in the complex $\omega$-plane is thus the straight horizontal line in the upper half-plane depicted in Fig.2(b). To ensure that $E_x(x, y, t)$ is real, it is necessary to have

$$\mathcal{E}_x(-k_x, -k_y, -\omega' + i\Omega) = \mathcal{E}_x^*(k_x, k_y, \omega' + i\Omega). \tag{4}$$

This is indeed the case, considering the definition of the Fourier-Laplace transform operator given by Eq.(2). [The general formula is $\mathcal{E}(-k_x^*, -k_y^*, -\omega^*) = \mathcal{E}^*(k_x, k_y, \omega)$ for any real-valued field $\boldsymbol{E}(x, y, z = 0, t)$.]

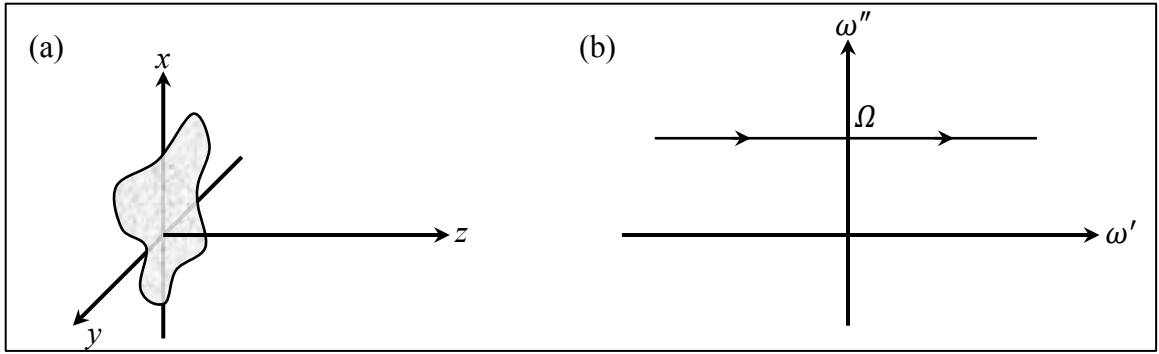

**Fig.2**. (a) In the Cartesian $xyz$ coordinate system, the predetermined $E$-field distribution in the $xy$-plane at $z = 0$ is given by Eq.(1). The initial state of the system at $t < 0$ is dark everywhere. The fields thus exist only for $t \geq 0$ and, in the absence of material media, the electromagnetic fields propagate in free space from left to right in the general direction of the $z$-axis. (b) In the complex $\omega$-plane, the inverse Laplace transform within Eq.(3) is carried out along the straight line $\omega = \omega' + i\Omega$ in the upper half-plane.

Maxwell's equations impose certain constraints on plane-waves propagating in free space. One constraint is that the $k$-vector and $\omega$ must satisfy the dispersion relation $\boldsymbol{k} \cdot \boldsymbol{k} = (\omega/c)^2$. Consequently,

$$\boxed{c = \text{speed of light in vacuum}} \rightarrow \quad k_z = \pm\sqrt{(\omega/c)^2 - k_x^2 - k_y^2}. \tag{5}$$

In choosing the plus or minus sign for $k_z$, one must remember that, in order to evaluate $E_x(x, y, z = z_0, t)$, the integrand in Eq.(3) must be multiplied by $\exp(ik_z z_0)$. As $\omega$ moves from left to right along the integration contour depicted in Fig.2(b), $k_x$ and $k_y$ in Eq.(5) remain fixed while $\omega'$ varies continuously from $-\infty$ to $+\infty$. The trajectory of the function $k_z^2 = (\omega'/c)^2 - [(\Omega/c)^2 + k_x^2 + k_y^2] + 2i\Omega\omega'/c^2$ in the complex-plane is the parabola shown in Fig.3, which crosses the real axis at a distance $(\Omega/c)^2 + k_x^2 + k_y^2$ to the left of the origin. It should not be difficult now to see that, of the two solutions for $k_z$, one is always in the upper-half of the complex plane, while the other is always in the lower-half.

For a beam propagating in vacuum along the positive $z$-axis, as depicted in Fig.2(a), the acceptable values of $k_z$ must be in the upper half-plane. Thus, as $\omega'$ moves from $-\infty$ to $+\infty$ along the integration path, the continuous variation of $k_z$ renders the function $\exp(ik_z z_0)$ analytic. Another way to observe this is to write

$$k_z = \left[(\omega/c) + \sqrt{k_x^2 + k_y^2}\right]^{\frac{1}{2}} \left[(\omega/c) - \sqrt{k_x^2 + k_y^2}\right]^{\frac{1}{2}}, \tag{6}$$

and to note that each of the two terms requires a branch-cut in the $\omega$-plane; see Appendix A for a brief discussion of such branch-cuts. Let the branch-cuts originate at $\omega = \pm c\sqrt{k_x^2 + k_y^2}$ and extend to $+\infty$ along the real axis. Thus, when $\omega'$ moves from $-\infty$ to $\infty$ along the integration path, the phase angles of both square roots in Eq.(6) will be in the $[0, \frac{1}{2}\pi]$ interval, ensuring that $k_z$ remains in the upper-half of the complex-plane. Moreover, since the branch-cuts are not crossed when $\omega$ moves along the integration path, the function $\exp(ik_z z_0)$ remains analytic along the entire path. (In fact, analyticity extends to the region $\omega'' \geq \Omega$, which is a requirement for causality.)

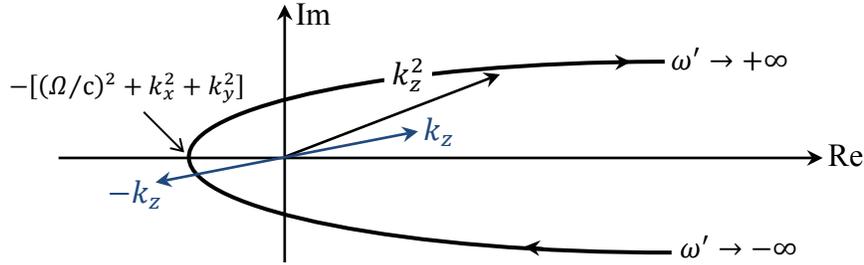

**Fig.3**. Trajectory of $k_z^2 = (\omega/c)^2 - k_x^2 - k_y^2$ for fixed values of $k_x$, $k_y$ and $\Omega$, as $\omega'$ moves from $-\infty$ to $+\infty$ along the straight line contour shown in Fig.2(b). For each value of $\omega'$, the correct $k_z$ is the one in the upper half-plane.

As an aside, note that the value of $k_z$ associated with $(-k_x, -k_y, -\omega' + i\Omega)$ is minus the conjugate of $k_z$ associated with $(k_x, k_y, \omega' + i\Omega)$. This is essential, since $\exp(ik_z z_0)$ must become its own complex-conjugate when the signs of $k_x$, $k_y$, and $\omega'$ are simultaneously reversed.

Another constraint imposed on plane-waves propagating in free space is that $\boldsymbol{k} \cdot \boldsymbol{E}_0$ must vanish. Considering that $\boldsymbol{k} \cdot \boldsymbol{E}_0 = k_x E_{x0} + k_y E_{y0} + k_z E_{z0}$, it is clear that the knowledge of $k_x$, $k_y$, $k_z$, $E_{x0}$, and $E_{y0}$ uniquely identifies $E_{z0}$. It is, therefore, not necessary to keep track of $E_z(\boldsymbol{r}, t)$, since the knowledge of $E_x$, $E_y$, and $\boldsymbol{k}$ is all that is needed to arrive at the $E_z$ distribution.

**Example**. Let $\boldsymbol{E}(x, y, z = 0, t) = E_{y0} \hat{\boldsymbol{y}} \, \text{step}(t) \cos(k_{x0} x - \omega_0 t)$, where $E_{y0}$, $k_{x0}$, and $\omega_0$ are positive real constants. The Fourier-Laplace transform of this $E$-field distribution at $k_x$, $k_y$, $\omega = \omega' + i\Omega$, where $\Omega > 0$, is readily found to be

$\mathcal{E}(k_x, k_y, \omega)$

$= \tfrac{1}{2} E_{y0} \hat{\boldsymbol{y}} \iint_{-\infty}^{\infty} \int_{t=0}^{\infty} \{\exp[i(k_{x0}x - \omega_0 t)] + \exp[-i(k_{x0}x - \omega_0 t)]\} \exp[-i(k_x x + k_y y - \omega t)] \, \mathrm{d}x \mathrm{d}y \mathrm{d}t$

$= \tfrac{1}{2} E_{y0} \hat{\boldsymbol{y}} \int_{-\infty}^{\infty} \exp(-ik_y y) \, \mathrm{d}y \, \{\int_{-\infty}^{\infty} \exp[-i(k_x - k_{x0})x] \mathrm{d}x \int_{t=0}^{\infty} \exp[i(\omega - \omega_0)t] \, \mathrm{d}t$

$\qquad + \int_{-\infty}^{\infty} \exp[-i(k_x + k_{x0})x] \mathrm{d}x \int_{t=0}^{\infty} \exp[i(\omega + \omega_0)t] \, \mathrm{d}t\}$

$= i2\pi^2 E_{y0} \hat{\boldsymbol{y}} \, \delta(k_y) \left[\dfrac{\delta(k_x - k_{x0})}{\omega - \omega_0} + \dfrac{\delta(k_x + k_{x0})}{\omega + \omega_0}\right]; \qquad (\omega = \omega' + i\Omega, \text{ where } \Omega > 0). \tag{7}$

Upon propagating in free space, at a distance $z = z_0 > 0$, the above transform will be multiplied by $\exp(ik_z z_0)$, where $k_z = \sqrt{(\omega/c)^2 - k_x^2 - k_y^2}$ is in the upper-half of the complex plane. An inverse Fourier-Laplace transformation then yields

$$\boldsymbol{E}(x, y, z_0, t)$$

$$= \frac{i2\pi^2 E_{y0}\hat{\boldsymbol{y}}}{(2\pi)^3} \iint_{-\infty}^{\infty} \int_{-\infty+i\Omega}^{\infty+i\Omega} \delta(k_y) \left[\frac{\delta(k_x - k_{x0})}{\omega - \omega_0} + \frac{\delta(k_x + k_{x0})}{\omega + \omega_0}\right] \exp(ik_z z_0) \exp[i(k_x x + k_y y - \omega t)] \, dk_x dk_y d\omega$$

$$= \frac{iE_{y0}\hat{\boldsymbol{y}}}{4\pi} \int_{-\infty+i\Omega}^{\infty+i\Omega} \left[\frac{\exp(ik_{x0} x)}{\omega - \omega_0} + \frac{\exp(-ik_{x0} x)}{\omega + \omega_0}\right] \exp\left[i\sqrt{(\omega/c)^2 - k_{x0}^2}\, z_0\right] \exp(-i\omega t) \, d\omega. \quad (8)$$

The above integral may, of course, be numerically evaluated along the specified path in the $\omega$-plane. Alternatively, one may invoke Cauchy's theorem of complex analysis to modify the path, and use the residue theorem to simplify the calculation.[4-6] Since the integrand is analytic above the horizontal line $\omega'' = \Omega$, we first consider a large semi-circle to close the integration path in the *upper-half* $\omega$-plane. On this semi-circle, $\omega \pm \omega_0 \to \omega$ and $k_z \to \omega/c$; therefore, the integrand in Eq.(8) approaches $\omega^{-1} \exp[i(\omega/c)(z_0 - ct)]$. Let $\omega = R\exp(i\varphi)$, where $R$ is the radius of the semi-circle and $\varphi$, the phase of the complex number $\omega$, is in the interval $(0, \pi)$. We may then write

$$|\omega^{-1} \exp[i(\omega/c)(z_0 - ct)]| = R^{-1} |\exp[i(R/c)(\cos\varphi + i\sin\varphi)(z_0 - ct)]|$$

$$= R^{-1} \exp[-(R/c) \sin\varphi\, (z_0 - ct)]. \quad (9)$$

Thus, for $z_0 > ct$, the integral over the upper semi-circle vanishes and, in the absence of poles and branch-cuts within this semi-circle, we conclude that the integral in Eq.(8) must vanish. This result should *not* be surprising, considering that $t$ is less than $z_0/c$, and that no signal can travel faster than the vacuum speed of light, $c$, in order to excite an electromagnetic field at $z = z_0$.

As for $t \geq z_0/c$, a closed contour in the lower half-plane is shown in Fig.4, where the contribution of the semi-circular segment to the closed-path integral vanishes when the radius of the semi-circle goes to infinity. The two poles at $\omega = \pm\omega_0$ then provide the steady-state ($ss$) solution at $z = z_0$, namely,

$$\boldsymbol{E}_{ss}(x, y, z_0, t) = E_{y0}\hat{\boldsymbol{y}} \cos\left[k_{x0} x + \sqrt{(\omega_0/c)^2 - k_{x0}^2}\, z_0 - \omega_0 t\right]. \quad (10)$$

The singularities of the integrand, however, are not confined to these two isolated poles because $\sqrt{(\omega/c)^2 - k_{x0}^2}$ is *not* an analytic function within the closed contour of integration. To see this, consider an analytic function $f(\omega)$, which has a zero at $\omega = \omega_0$ and a nonzero derivative at that point, that is, $f(\omega_0) = 0$ and $f'(\omega_0) \neq 0$. For $\omega$ sufficiently close to $\omega_0$ we may write

$$f'(\omega_0) \cong \frac{f(\omega) - f(\omega_0)}{\omega - \omega_0} \quad \to \quad f(\omega) \cong f'(\omega_0)(\omega - \omega_0). \quad (11)$$

Thus, as $\omega$ traverses a full circle around $\omega_0$, $f(\omega)$ undergoes a $2\pi$ phase change. Consequently, the phase of $\sqrt{f(\omega)}$ changes by $\pi$ around the circle, creating a discontinuity that requires a branch-cut to render $\sqrt{f(\omega)}$ analytic everywhere in the vicinity of $\omega_0$ except on the branch-cut. The branch-cut originates at $\omega_0$, but may have arbitrary shape and direction otherwise (see Appendix A). Since $k_z$ has two zeros at $\omega = \pm ck_{x0}$, the branch-cut in Fig.4 is chosen to be the straight-line connecting these two points in the $\omega$-plane. The transient solution for $t \geq z_0/c$ is now obtained by integrating the integrand of Eq.(8) around the branch-cut, namely,

$$E_{\text{transient}}(x, y, z_0, t \geq z_0/c)$$

$$= \frac{iE_{y0}\hat{y}}{\pi} \int_{-ck_{x0}}^{ck_{x0}} \left[\frac{\omega \cos(k_{x0}x) + i\omega_0 \sin(k_{x0}x)}{\omega_0^2 - \omega^2}\right] \sinh\left[\sqrt{(ck_{x0})^2 - \omega^2}\,(z_0/c)\right] \exp(-i\omega t)\, d\omega$$

$$= \frac{2E_{y0}\hat{y}}{\pi} \int_{\omega=0}^{ck_{x0}} \left[\frac{\omega \sin(\omega t)\cos(k_{x0}x) - \omega_0 \cos(\omega t)\sin(k_{x0}x)}{\omega_0^2 - \omega^2}\right] \sinh\left[\sqrt{(ck_{x0})^2 - \omega^2}\,(z_0/c)\right] d\omega. \qquad (12)$$

As expected, the transient solution vanishes for $t \gg z_0/c$, because of the rapid oscillations of $\sin(\omega t)$ and $\cos(\omega t)$ appearing in the above equation.

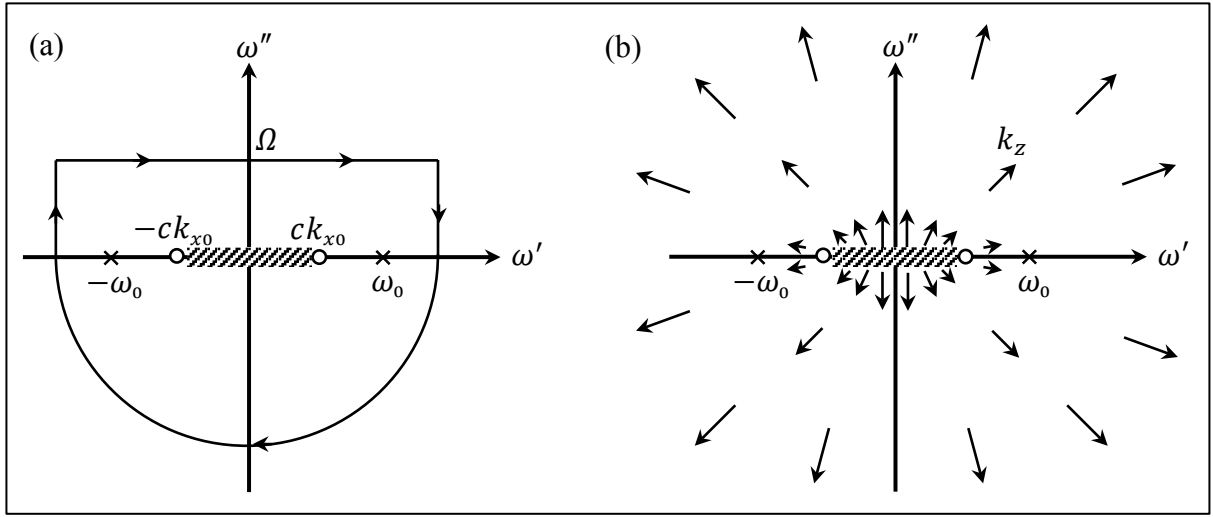

**Fig.4**. (a) The closed contour of integration in the $\omega$-plane consists of the straight horizontal line $\omega = \omega' + i\Omega$, two short vertical legs at $\omega' = \pm\infty$, and an infinitely large semi-circle in the lower-half of the complex plane. The integrand in Eq.(8) has two simple poles at $\omega = \pm\omega_0$. The two zeros of $k_z$ located at $\omega = \pm ck_{x0}$ define a branch-cut (cross-hatched). (b) The choice of sign for $k_z = \pm\sqrt{(\omega/c)^2 - k_{x0}^2}$ ensures that the function $\exp(ik_z z_0)$ is analytic throughout the $\omega$-plane except at the branch-cut located between the two zeros of $k_z$. Far away from the origin, $k_z \cong \omega/c$. Note the discontinuity of $k_z$ above and below the branch-cut, and also its 180° rotation as $\omega$ traverses a full circle around each zero of $k_z$. On the straight horizontal line $\omega = \omega' + i\Omega$, where the integral in Eq.(8) is to be evaluated, the chosen sign for $k_z$ ensures its mandatory confinement to the upper-half of the complex plane.

**3. Poles, zeros, and branch-cuts of $k_z$.** A plane-wave propagating in a homogeneous, isotropic, linear medium is specified by a complex frequency $\omega$ and a complex $k$-vector $\boldsymbol{k} = k_x\hat{x} + k_y\hat{y} + k_z\hat{z}$, where, in general, $k_x, k_y, k_z$ are complex-valued. Maxwell's equations then require that $k_x^2 + k_y^2 + k_z^2 = \mu(\omega)\varepsilon(\omega)(\omega/c)^2$, where $\mu(\omega)$ is the relative permeability and $\varepsilon(\omega)$ the relative permittivity of the host medium. For a non-magnetic medium supported by a single Lorentz oscillator, we have $\mu(\omega) = 1$ and $\varepsilon(\omega) = 1 \pm \frac{\omega_p^2}{\omega_r^2 - \omega^2 - i\gamma\omega}$, where $\omega_p$ is the plasma frequency, $\omega_r$ is the resonance frequency, $\gamma$ (a positive real constant) is the loss/gain coefficient, and the plus (minus) sign applies to a lossy (gainy) medium. In what follows, we shall assume that $k_y = 0$, and proceed to write

$$k_z = \sqrt{\mu(\omega)\varepsilon(\omega)(\omega/c)^2 - k_x^2} = \left\{\frac{[\omega^2 + i\gamma\omega - (\omega_r^2 \pm \omega_p^2)](\omega/c)^2}{\omega^2 + i\gamma\omega - \omega_r^2} - k_x^2\right\}^{1/2} \qquad \text{}$$

$$= \left[ \frac{\left(\omega+\tfrac{1}{2}i\gamma-\sqrt{\omega_r^2\pm\omega_p^2-\tfrac{1}{4}\gamma^2}\right)\left(\omega+\tfrac{1}{2}i\gamma+\sqrt{\omega_r^2\pm\omega_p^2-\tfrac{1}{4}\gamma^2}\right)\left(\tfrac{\omega}{c}\right)^2}{\left(\omega+\tfrac{1}{2}i\gamma-\sqrt{\omega_r^2-\tfrac{1}{4}\gamma^2}\right)\left(\omega+\tfrac{1}{2}i\gamma+\sqrt{\omega_r^2-\tfrac{1}{4}\gamma^2}\right)} - k_x^2 \right]^{1/2} = \left[ \frac{(\omega-\omega_1)(\omega-\omega_2)(\omega/c)^2}{(\omega-\omega_3)(\omega-\omega_4)} - k_x^2 \right]^{1/2}. \quad (13)$$

For weakly lossy/gainy media, $\gamma$ is small, and we can safely assume that $\omega_r \gg \gamma$. The poles of $k_z$ are always going to be in the lower-half of the $\omega$-plane at $\omega = \omega_3$ and $\omega = \omega_4$, independently of the value of $k_x$, and irrespective of whether the medium is lossy or gainy. As for the zeros of $k_z$, we note that, for any given $\omega$, there will be two points in the complex-plane of $k_x$, namely,

$$k_{x0} = \pm\left(\tfrac{\omega}{c}\right)\sqrt{\frac{(\omega-\omega_1)(\omega-\omega_2)}{(\omega-\omega_3)(\omega-\omega_4)}}, \quad (14)$$

at which $k_z$ vanishes. To simplify the forthcoming analysis, we shall assume that, in the case of gainy media, $\omega_r^2 > (\omega_p^2 + \tfrac{1}{4}\gamma^2)$, so that both $\omega_1$ and $\omega_2$ are in the lower-half of the $\omega$-plane, located on the horizontal line connecting $\omega_3$ to $\omega_4$. (Appendix B examines the $\omega$-plane trajectories of the zeros of $k_z$ when $k_x$ moves along the real-axis $k_x'$ of the $k_x$-plane.)

Our next task is to fix the real part of $\omega = \omega' + i\omega''$ on the real-axis of the $\omega$-plane, then move up parallel to the imaginary axis (i.e., move $\omega''$ from 0 to $+\infty$), and identify the corresponding trajectories of the two zeros of $k_z$, given by Eq.(14), within the complex $k_x$-plane. Considering that $\omega_1'' = \omega_2'' = \omega_3'' = \omega_4'' = -\gamma/2$, we may define $\varphi_{1,2}$ as the sum of the angles of $\omega - \omega_1$ and $\omega - \omega_2$, as follows:

$$\varphi_{1,2} = \tan^{-1}\left(\frac{\omega''+\tfrac{1}{2}\gamma}{\omega'-\omega_1'}\right) + \tan^{-1}\left(\frac{\omega''+\tfrac{1}{2}\gamma}{\omega'-\omega_2'}\right) \rightarrow \varphi_{1,2} = \tan^{-1}\left[\frac{2\omega'(\omega''+\tfrac{1}{2}\gamma)}{\omega'^2-(\omega''+\tfrac{1}{2}\gamma)^2-\omega_1'^2}\right]. \quad (15a)$$

Similarly, the sum of the angles of $\omega - \omega_3$ and $\omega - \omega_4$ is given by

$$\varphi_{3,4} = \tan^{-1}\left(\frac{\omega''+\tfrac{1}{2}\gamma}{\omega'-\omega_3'}\right) + \tan^{-1}\left(\frac{\omega''+\tfrac{1}{2}\gamma}{\omega'-\omega_4'}\right) \rightarrow \varphi_{3,4} = \tan^{-1}\left[\frac{2\omega'(\omega''+\tfrac{1}{2}\gamma)}{\omega'^2-(\omega''+\tfrac{1}{2}\gamma)^2-\omega_3'^2}\right]. \quad (15b)$$

Thus, at $\omega = \omega' + i\omega''$, where $\omega'' \geq 0$, the phase-angle of $k_{x0}$ (a zero of $k_z$) is given by

$$\varphi_{k_{x0}} = \tan^{-1}(\omega''/\omega') + \tfrac{1}{2}(\varphi_{1,2} - \varphi_{3,4}). \quad (15c)$$

In Eqs.(15), the arctangent must be understood as the angle residing in the $[0, \pi]$ interval when $\omega' \geq 0$. The same holds true for $\omega' < 0$, except that, in the second halves of Eqs.(15a) and (15b), the arctangent signifies an angle belonging to the $[\pi, 2\pi]$ interval. While the detailed trajectory of $k_{x0}$ in the $k_x$-plane, as $\omega''$ moves from 0 to $+\infty$, depends on the value of $\omega'$ as well as on the parameters of the medium (i.e., $\omega_p, \omega_r, \gamma$), the general behavior can be seen from Eq.(15c) to be as depicted in Fig.5. The trajectories of $k_{x0}$ for a lossy medium ($\omega_1' > \omega_3'$) are shown in Fig.5(a), while those for a gainy medium ($\omega_1' < \omega_3'$) appear in Fig.5(b). The main difference between the two cases is that the trajectories $\pm k_{x0}(\omega'')$ cross the real axis when the medium is gainy, but do not do so in the case of lossy media. Although Fig.5 shows the trajectories of the two zeros of $k_z$ in the $k_x$-plane for a typical value of $\omega' > 0$, it is not difficult to see from Eq.(14) that replacing $\omega$ with $-\omega^*$ (i.e., changing the sign of $\omega'$) will cause $k_{x0}$ to change to $-k_{x0}^*$. [The same conclusion can be reached by examining Eqs.(15).]

Note that, in accordance with the first line of Eq.(13), the starting points of the two trajectories in the $k_x$-plane are located at

$$k_{x0}(0) = \pm\sqrt{\mu(\omega')\varepsilon(\omega')}\,(\omega'/c) = \pm n(\omega')\,\omega'/c. \quad (16)$$

Considering that typical lossy media have a refractive index $n(\omega')$ located in the first quadrant ($Q_1$) of the complex-plane, whereas the refractive index of a typical gainy medium is in the fourth quadrant ($Q_4$), it is now easy to identify with $n(\omega')$ the location of the starting points of the trajectories $k_{x0}(\omega'')$ depicted in Fig.5.

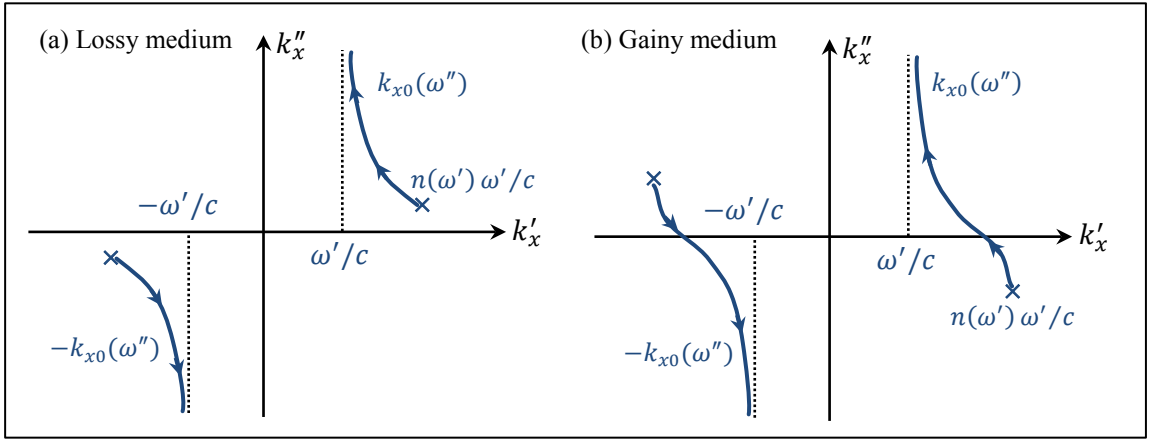

**Fig.5**. Trajectories of the two zeros of $k_z$ in the complex $k_x$-plane, when the real part of the frequency $\omega = \omega' + i\omega''$ is kept constant while its imaginary part rises from 0 to ∞ in the upper half of the $\omega$-plane. (a) The case of a lossy medium, where $\omega_1' > \omega_3'$. (b) The case of a gainy medium, where $\omega_1' < \omega_3'$. In both cases, $\omega'$ is taken to be positive. (For $\omega' < 0$, the above trajectories would flip around the vertical axis.) Note, in the case of a lossy medium, that the trajectory of $k_{x0}(\omega'')$ does *not* cross the real axis.

The above analysis shows that a decomposition of the incident beam via a Fourier transform along the $x$-axis will result, in the case of gainy media, in certain values of the real Fourier variable $k_x$ being associated with a zero of $k_z$ in the upper-half $\omega$-plane. If, however, we were to *deform* the contour of integration in the $k_x$-plane in such a way as to avoid running into the $\pm k_{x0}(\omega'')$ trajectories, as shown in Fig.6, then, in accordance with Cauchy's theorem,[4-6] the contour of the inverse Laplace integral in the $\omega$-plane could be lowered to coincide with the real-axis $\omega'$. (Appendix C describes the Fourier-Laplace transform and the mathematics of deformed contours.) In other words, the inverse Laplace integral over $\omega = \omega' + i\Omega$ would become an inverse Fourier integral over $\omega'$. One may now write, for any given value of $\omega'$,

$$k_z(k_x, \omega') = \pm i \sqrt{[k_x + n(\omega')\omega'/c][k_x - n(\omega')\omega'/c]}; \quad (\pm \text{ sign for } \omega' \lessgtr 0), \quad (17)$$

where the plus sign is used for $\omega' < 0$, and the minus sign for $\omega' > 0$, so that $k_z(-k_x^*, -\omega^*) = -k_z^*(k_x, \omega)$, as it must. It should be remembered that each of the two terms under the radical in Eq.(17) is associated with a branch-cut in the $k_x$-plane. In fact, a good choice for these branch-cuts would be the (meandering) trajectories $\pm k_{x0}(\omega'')$ themselves, as illustrated in Fig.6.

In writing the expression for $k_z$ in Eq.(17), we have factored out $\sqrt{-1}$ and written it as $\pm i$, with the plus and minus signs associated with $\omega' < 0$ and $\omega' > 0$, respectively. This convenient choice of signs will simplify later analysis, considering that a plane-wave's phase-factor, $\exp(ik_z z)$, may now be written as $\exp[\sqrt{(k_x + k_{x0})(k_x - k_{x0})}\, z]$ in the case of $\omega' > 0$. Figures 7(a) and 7(b) illustrate, for lossy and gainy media respectively, the method of calculation of $k_z$ at a given frequency $\omega'$ and for different values of $k_x$. Special attention must be paid to the branch-cuts in order to determine the correct phase-angle for $k_z$ at each point.

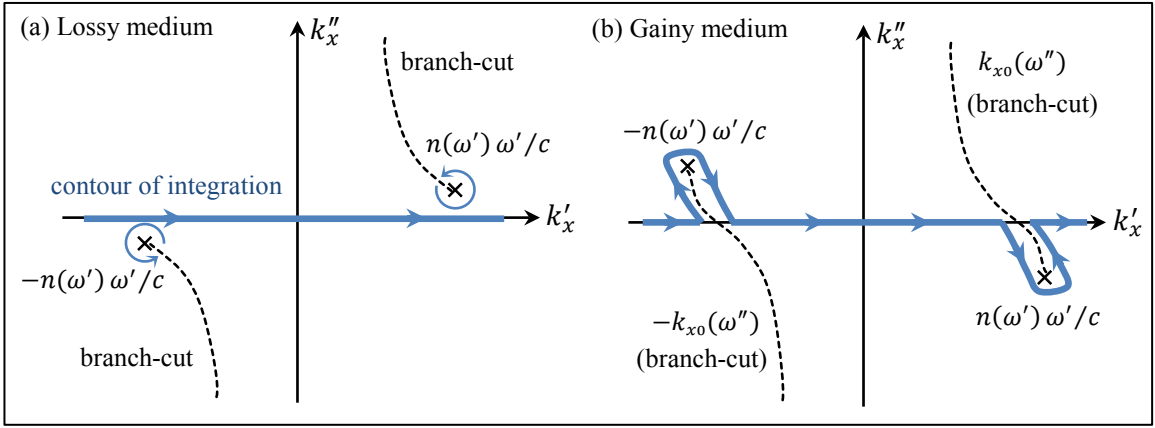

**Fig.6**. Contour of integration in the complex $k_x$-plane. The trajectories of $k_{x0}(\omega'')$ now serve as branch-cuts. (a) In the case of lossy media, the integration contour is the same as the real axis $k_x'$. The circles surrounding the points $k_{x0}(0) = \pm n(\omega')\omega'/c$ indicate the angular range for each branch-cut. (b) In the case of gainy media, the contour must be deformed to avoid any overlaps with $k_{x0}(\omega'')$. Since the branch-cuts are chosen to coincide with the $k_{x0}(\omega'')$ trajectories, the deformed contour may closely hug the branch-cut. Both diagrams in (a) and (b) correspond to $\omega' > 0$. For $\omega' < 0$, the $k_{x0}(\omega'')$ trajectories, which serve here as branch-cuts, would be flipped around the vertical axis.

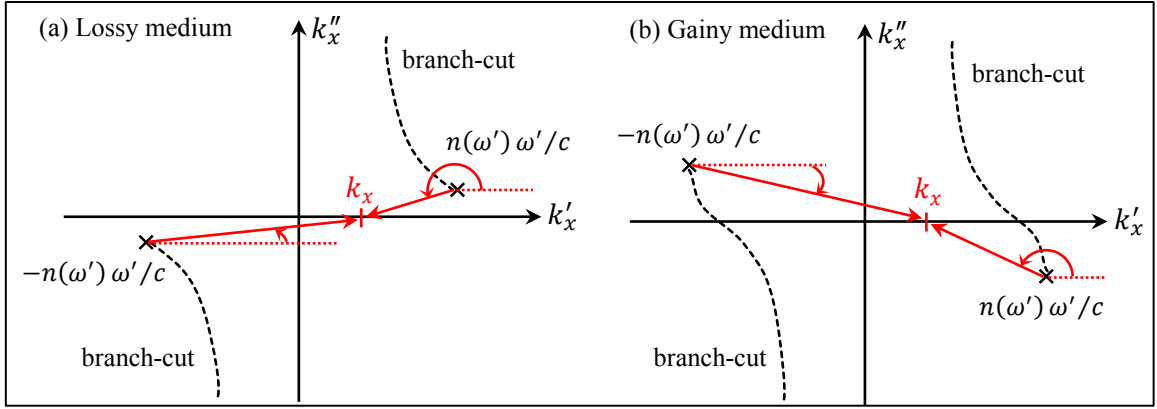

**Fig.7**. To evaluate $k_z$ in accordance with Eq.(17), one must determine the phase-angles of the two complex numbers obtained by connecting the zeros of $k_z$ to each point $k_x$ on the contour of integration. The acceptable value for each angle is intimately tied to the corresponding branch-cut. In (a), which represents a lossy medium at $\omega' > 0$, the angle of the arrow on the left is slightly larger than zero, while that of the arrow on the right is somewhat greater than 180°. Thus the radical in Eq.(17), which is equal to $ik_z$ for $\omega' > 0$, has an angle somewhat greater than 90°, indicating that $\exp(ik_z z)$ propagates forward along the $z$-axis, while slowly decaying. If $k_x$ moves to the far right of the $k_x'$-axis on the diagram in (a), the two arrows will have angles slightly greater than zero and somewhat less than 360°, respectively. The phase-angle of the radical (i.e., $ik_z$) will then be a little less than 180°, making $\exp(ik_z z)$ decay rapidly along the $z$-axis while having a positive phase velocity. In (b), which represents the case of a gainy medium at $\omega' > 0$, the angle of one arrow is slightly less than zero, while that of the other is somewhat below 180°. The phase of the radical, therefore, is a little below 90°, consistent with $\exp(ik_z z)$ propagating forward along the $z$-axis, while slowly growing in magnitude. Finally, if $k_x$ moves to the far right on the $k_x'$-axis of the diagram in (b), one arrow's angle will be slightly below zero while that of the other one will be somewhat greater than 360°. The angle of the radical will then be just over 180°, making $\exp(ik_z z)$ decay rapidly along the $z$-axis while propagating backward, having a negative phase-velocity.

**Example**. Consider the following $E$-field distribution in the $xy$-plane at $z = 0$:

$$\boldsymbol{E}(x, y, z = 0, t) = E_{y0}\hat{\boldsymbol{y}} \exp(-x^2/w^2) \cos(\kappa x - \omega_c t)\text{step}(t). \tag{18}$$

Here, $E_{y0}$ is the field amplitude at $x = 0$, $w$ is the width of the Gaussian beam profile along the $x$-axis, $\omega_c$ is the central oscillation frequency, and the real-valued $\kappa$ is related to the tilt angle of the beam in the $xz$-plane away from the $z$-axis. Expressing the above initial distribution as a superposition of plane-waves (having complex $\omega$) requires a Fourier-Laplace transform, as follows:

$$\mathcal{E}_y(k_x, k_y, \omega) = \iint_{x,y=-\infty}^{\infty} \int_{t=0}^{\infty} E_y(x, y, z = 0, t) \exp[-\mathrm{i}(k_x x + k_y y - \omega t)]\, \mathrm{d}x\mathrm{d}y\mathrm{d}t$$

$$= \pi E_{y0}\delta(k_y)\{\int_{-\infty}^{\infty} \exp(-x^2/w^2)\exp[-\mathrm{i}(k_x - \kappa)x]\, \mathrm{d}x \int_0^{\infty} \exp[\mathrm{i}(\omega - \omega_c)t]\, \mathrm{d}t$$

$$+ \int_{-\infty}^{\infty} \exp(-x^2/w^2)\exp[-\mathrm{i}(k_x + \kappa)x]\, \mathrm{d}x \int_0^{\infty} \exp[\mathrm{i}(\omega + \omega_c)t]\, \mathrm{d}t\}$$

$$= \mathrm{i}\pi\sqrt{\pi}wE_{y0}\delta(k_y)\left\{\frac{\exp[-\tfrac{1}{4}w^2(k_x - \kappa)^2]}{\omega - \omega_c} + \frac{\exp[-\tfrac{1}{4}w^2(k_x + \kappa)^2]}{\omega + \omega_c}\right\}. \tag{19}$$

Upon propagating a distance $z = z_0 > 0$ within a homogeneous, isotropic, non-magnetic, linear medium having dielectric function $\varepsilon(\omega)$, the above Fourier-Laplace transform will be multiplied by $\exp(\mathrm{i}k_z z_0)$, where $k_z = \sqrt{\varepsilon(\omega)(\omega/c)^2 - k_x^2 - k_y^2}$, albeit with proper attention paid to its sign. An inverse Fourier-Laplace transformation then yields

$$E_y(x, y, z_0, t) = \frac{\mathrm{i}\pi\sqrt{\pi}wE_{y0}}{(2\pi)^3} \int_{-\infty+\mathrm{i}\Omega}^{\infty+\mathrm{i}\Omega} \iint_{-\infty}^{\infty} \delta(k_y) \left\{\frac{\exp[-\tfrac{1}{4}w^2(k_x - \kappa)^2]}{\omega - \omega_c} + \frac{\exp[-\tfrac{1}{4}w^2(k_x + \kappa)^2]}{\omega + \omega_c}\right\} \exp(\mathrm{i}k_z z_0)$$

$$\times \exp[\mathrm{i}(k_x x + k_y y - \omega t)]\, \mathrm{d}k_x \mathrm{d}k_y \mathrm{d}\omega$$

$$= \frac{\mathrm{i}wE_{y0}}{8\pi\sqrt{\pi}} \int_{-\infty+\mathrm{i}\Omega}^{\infty+\mathrm{i}\Omega} \exp(-\mathrm{i}\omega t) \int_{-\infty}^{\infty} \left\{\frac{\exp[-\tfrac{1}{4}w^2(k_x - \kappa)^2]}{\omega - \omega_c} + \frac{\exp[-\tfrac{1}{4}w^2(k_x + \kappa)^2]}{\omega + \omega_c}\right\}$$

$$\times \exp\!\left[\pm\mathrm{i}\sqrt{\varepsilon(\omega)(\omega/c)^2 - k_x^2}\, z_0\right] \exp(\mathrm{i}k_x x)\, \mathrm{d}k_x \mathrm{d}\omega. \tag{20}$$

For each $\omega = \omega'$ on the real-axis of the $\omega$-plane, and for given values of $x$ and $z_0$, the integral over $k_x$ in Eq.(20) may now be evaluated on a deformed path in the $k_x$-plane, as described in the preceding section. Consequently, the contour of integration in the $\omega$-plane can be brought down to coincide with the real-axis $\omega'$, except in the vicinity of $\omega = \pm \omega_c$, where the contour must be bent upward (ever so slightly) to avoid contacting these two poles of the integrand. The integral over $k_x$, however, will not be affected by this slight upward bend of the contour in the $\omega$-plane (around $\omega = \pm \omega_c$), because the integrand of the $k_x$ integral is a continuous function of $\omega$ at $\omega = \pm \omega_c$.

In the steady state, when $t \to \infty$, the contour of integration in the $\omega$-plane may be closed with a large semi-circle in the lower half-plane. Any poles or branch-cuts located in the lower half-plane will not contribute to the steady-state ($ss$) solution, because their residues will be proportional to $\exp(\omega'' t)$ where $\omega'' < 0$ and $t \to \infty$. These residues, therefore, will vanish in the steady state. The only remaining contributions come from the two poles at $\omega = \pm \omega_c$. We will have

$$E_y^{(ss)}(x, y, z_0, t)$$

$$= \frac{wE_{y0}}{4\sqrt{\pi}}\left\{\exp(-\mathrm{i}\omega_c t) \int_{\text{path}(\omega_c)} \exp[-\tfrac{1}{4}w^2(k_x - \kappa)^2] \exp\!\left[\mathrm{i}k_x x + \sqrt{k_x^2 - \varepsilon(\omega_c)(\omega_c/c)^2}\, z_0\right] \mathrm{d}k_x\right.$$

$$+ \exp(i\omega_c t) \int_{\text{path}(-\omega_c)} \exp[-\tfrac{1}{4}w^2(k_x + \kappa)^2] \exp\left[ik_x x - \sqrt{k_x^2 - \varepsilon(-\omega_c)(\omega_c/c)^2}\, z_0\right] dk_x \Big\}$$

$$= \frac{wE_{y0}}{2\sqrt{\pi}} \text{Real}\left\{\exp(-i\omega_c t) \int_{\text{path}(\omega_c)} \exp[-\tfrac{1}{4}w^2(k_x - \kappa)^2] \exp\left[ik_x x + \sqrt{k_x^2 - \varepsilon(\omega_c)(\omega_c/c)^2}\, z_0\right] dk_x\right\}. \quad (21)$$

Note that the second integral in Eq.(21) is the conjugate of the first, as may be readily verified by changing the variable from $k_x$ to $-k_x^*$, noting that the $k_x$-plane integration paths corresponding to $\omega_c$ and $-\omega_c$ are mirror images of each other (i.e., flipped around the imaginary $k_x''$-axis).

**4. The causality constraint**. The $E$-field distribution at a distance $z = z_0$ from the entrance facet of a semi-infinite medium occupying the half-space $z \geq 0$ is given by

$$E(x, z_0, t) = (2\pi)^{-2} \iint_{-\infty}^{\infty} \mathcal{E}(k_x, \omega) \exp\{i[k_x x + k_z z_0 - (\omega' + i\Omega)t]\} \, dk_x d\omega'. \quad (22)$$

Here, the dispersion relation $\boldsymbol{k} \cdot \boldsymbol{k} = (\omega/c)^2 \mu(\omega)\varepsilon(\omega)$ yields

$$k_z(k_x, \omega) = \pm\sqrt{(\omega/c)^2 \mu(\omega)\varepsilon(\omega) - k_x^2}. \quad (23)$$

The sign of $k_z$ must be chosen such that, when $t < z_0/c$, the integral in Eq.(22) will vanish.[2,3] Upon closing the integration path in the $\omega$-plane via a large semi-circle in the upper half-plane, given that $\mu(\omega) \to 1$ and $\varepsilon(\omega) \to 1$ everywhere on the large semi-circle when $|\omega| \to \infty$, one readily concludes that choosing the plus sign for the square root in Eq.(23) ensures the vanishing of the integral on the infinitely large semi-circle. The choice of branch-cuts for $k_z$ is thus essential in the context of causality, as they are intimately tied to the analyticity of $k_z$ in the region above the straight line $\omega = \omega' + i\Omega$ as well as its asymptotic behavior when $|\omega| \to \infty$ on the large semi-circle that closes the integration path in the upper half of the $\omega$-plane.

**5. The incident wave-packet**. The excitation must start at a finite instant in time, say, at $t = t_0$. That way, when the electromagnetic field grows in time, assuming the growth rate does not exceed $e^{\Omega t}$ for some fixed, positive $\Omega$, we can multiply the field everywhere, i.e., at all points $(x, y, z)$ within the system of interest, by $e^{-\Omega t}$, then Fourier transform the resulting field. The $E$-field then becomes a superposition of plane-waves $\mathcal{E}(\boldsymbol{k}, \omega) \exp\{i[\boldsymbol{k} \cdot \boldsymbol{r} - (\omega' + i\Omega)t]\}$, where $\boldsymbol{k} = (k_x, k_y, k_z)$ is the wave-vector within the medium that is host to the electromagnetic field. Thus, in the $\omega$-plane, the inverse Fourier integral is not taken along the real axis, $\omega'$, but rather along a straight line parallel to $\omega'$ at a distance of $\omega'' = \Omega$. (See Appendix D for a fundamental property of functions that have either a finite width — in space or in time — or a finite bandwidth in their respective Fourier domain. Appendix E describes certain important characteristics of functions of $x$ and $t$ that have a finite width in both dimensions.)

Let the incident wave-packet have a finite width $w$ along the $x$-axis and a finite duration $\tau$ in time. Moreover, we assume that the incident packet has a center frequency $\omega_c$ as well as a tilt toward the $x$-axis, represented by the fixed parameter $k_{xc}$. (Here, $w$, $\tau$, $\omega_c$ and $k_{xc}$ are real-valued.) The general spatio-temporal dependence of the packet is specified as

$$E^{(\text{inc})}(x, t) = f(x/w)g(t/\tau)\cos(k_{xc}x - \omega_c t). \quad (24)$$

The choice of the functions $f(\cdot)$ and $g(\cdot)$ is more or less arbitrary, so long as they vanish outside a certain range of their respective variables. Nevertheless, it is preferable if these functions have a narrowband Fourier spectrum that would rapidly decline outside their nominal bandwidth. A reasonable choice for both functions is $\text{rect}(\cdot) * \text{rect}(\cdot) * \cdots * \text{rect}(\cdot)$, where the $\text{rect}(\cdot)$ function equals 1.0 when its argument is between $-\tfrac{1}{2}$ and $\tfrac{1}{2}$, and 0 otherwise. The number of times $\text{rect}(\cdot)$ is convolved with itself is arbitrary, but should be kept reasonably small, say, 2, or 3, or 4 times. Also, it is not necessary for the number of convolutions used for $f(\cdot)$ to be the same as that for $g(\cdot)$.

The Fourier transform of rect($x$) is $\int_{-½}^{½} e^{-ik_x x} dx = \sin(k_x/2)/(k_x/2) = \text{sinc}(k_x/2\pi)$.[†] Thus, with $n$ convolutions, the Fourier transform of $f(x/w)$ will be $\tilde{f}(k_x) = w\,\text{sinc}^n(wk_x/2\pi)$, while that of $g(t/\tau)$, following $m$ convolutions, will be $\tilde{g}(\omega) = \tau\,\text{sinc}^m(\tau\omega/2\pi)$. Therefore, the Fourier transform of the incident $E$-field of Eq.(24) is given by

$$\mathcal{E}^{(\text{inc})}(k_x, \omega) = ½\tilde{f}(k_x - k_{xc})\tilde{g}(\omega - \omega_c) + ½\tilde{f}(k_x + k_{xc})\tilde{g}(\omega + \omega_c). \tag{25}$$

A reasonable set of parameter values that has been used in our numerical work[1] is $\omega_c = 2.5 \times 10^{15}$ rad/s (corresponding to the vacuum wavelength $\lambda_c = 0.754$ μm), $k_{xc} = 8 \times 10^6$ m$^{-1}$ (resulting in an incidence angle of ~74° in free space), $\tau = 10^{-14}$ sec, and $w = 5 \times 10^{-6}$ m.

**6. Reflection and transmission at the interface between a dielectric host and a semi-infinite gain medium**. The single-surface problem depicted in Fig.1(a) is, in some ways, simpler than the finite-thickness slab problem of Fig.1(b). This is because the single-surface transfer functions (i.e., the Fresnel reflection and transmission coefficients) have no poles in the upper-half $\omega$-plane. The sole reason for lack of analyticity in the upper-half $\omega$-plane is the existence in this region of a branch-cut of $k_{2z}$, which necessitates a shift from an ordinary Fourier transformation on the real $\omega'$-axis to a Fourier-Laplace transform on the straight-line $\omega = \omega' + i\Omega$, where $\Omega > 0$ places the straight-line above the troublesome branch-cut. We will then have the option of (i) using a conventional Fourier transform on the real $k'_x$-axis, followed by a Fourier-Laplace transform on the straight line located at $\omega'' = \Omega$, or (ii) switch to a deformed integration contour in the $k_x$-plane that would remove the branch-cut from the upper-half $\omega$-plane, followed by an ordinary Fourier transformation on the $\omega'$-axis. The two methods are expected to yield identical results in the end.

For the sake of concreteness, let us fix the parameters of the incidence medium 1 at $\omega_{p1} = 5$, $\omega_{r1} = 15$, $\gamma_1 = 1$, and those of the gain medium 2 at $\omega_{p2} = 1.1$, $\omega_{r2} = 10$, $\gamma_2 = 1$. (These parameters, which are normalized by $\omega_0 = 3 \times 10^{14}$ rad/sec, have been used in our numerical work reported in [1].) As usual, the single-Lorentz-oscillator model of the dielectric functions yields

$$\varepsilon_1(\omega) = 1 + \frac{\omega_{p1}^2}{\omega_{r1}^2 - \omega^2 - i\gamma_1\omega}, \tag{26}$$

$$\varepsilon_2(\omega) = 1 - \frac{\omega_{p2}^2}{\omega_{r2}^2 - \omega^2 - i\gamma_2\omega}. \tag{27}$$

Each medium will have its own $k_z$, which is a function of $k_x$, $\omega$, and the corresponding $\varepsilon(\omega)$, as follows:

$$k_z = (\omega/c)\sqrt{\varepsilon(\omega) - (ck_x/\omega)^2}. \tag{28}$$

The signs of $k_{1z}$ and $k_{2z}$ at any given point $(k_x, \omega)$ are uniquely specified by the requirement of causality,[2,3] which states that, for an incident wave-packet arriving at $t = 0$ at the interface (i.e., at $z = 0$), no signal can reach the point $(x, y, z)$ prior to $t = |z|/c$. Enforcement of the causality requirement for the reflected wave locks-in the sign of $k_{1z}$, while its enforcement for the transmitted wave uniquely specifies the sign of $k_{2z}$.

---

[†]Here, we are using Bracewell's definition of the function sinc($x$) as $\sin(\pi x)/(\pi x)$.[9]

For an $s$-polarized incident wave-packet (i.e., one whose $E$-field is aligned with the $y$-axis), the Fresnel reflection and transmission coefficients are readily found to be[2,3,8]

$$\rho(k_x, \omega) = (k_{1z} - k_{2z})/(k_{1z} + k_{2z}), \tag{29}$$

$$\tau(k_x, \omega) = 2k_{1z}/(k_{1z} + k_{2z}). \tag{30}$$

It is not difficult to confirm that $\rho$ and $\tau$ share a pair of simple poles in the lower-half $\omega$-plane, independently of the specific value of $k_x$, and that these poles pose no difficulties for our analysis of the reflection/transmission processes.

---

Specifically, the pair of poles shared by $\rho$ and $\tau$ are evaluated as follows:

$$k_{1z} + k_{2z} = 0 \quad \rightarrow \quad (\omega/c)\sqrt{\varepsilon_1(\omega) - (ck_x/\omega)^2} \pm (\omega/c)\sqrt{\varepsilon_2(\omega) - (ck_x/\omega)^2} = 0$$

$$\rightarrow \quad \varepsilon_1(\omega) = \varepsilon_2(\omega) \quad \rightarrow \quad \frac{\omega_{p1}^2}{\omega_{r1}^2 - \omega^2 - i\gamma_1\omega} = -\frac{\omega_{p2}^2}{\omega_{r2}^2 - \omega^2 - i\gamma_2\omega}$$

$$\rightarrow \quad (\omega_{p1}^2 + \omega_{p2}^2)\omega^2 + i(\gamma_1\omega_{p2}^2 + \gamma_2\omega_{p1}^2)\omega - (\omega_{r1}^2\omega_{p2}^2 + \omega_{r2}^2\omega_{p1}^2) = 0$$

$$\rightarrow \quad \omega = \pm\sqrt{\frac{\omega_{p1}^2\omega_{r2}^2 + \omega_{p2}^2\omega_{r1}^2}{\omega_{p1}^2 + \omega_{p2}^2} - \tfrac{1}{4}\left(\frac{\gamma_1\omega_{p2}^2 + \gamma_2\omega_{p1}^2}{\omega_{p1}^2 + \omega_{p2}^2}\right)^2} - \tfrac{1}{2}i\left(\frac{\gamma_1\omega_{p2}^2 + \gamma_2\omega_{p1}^2}{\omega_{p1}^2 + \omega_{p2}^2}\right). \tag{31}$$

Given our usual assumptions about the magnitudes of the material parameters, both poles identified by Eq.(31) should reside in the lower-half of the $\omega$-plane. Note that, these poles may or may not exist, depending on the signs chosen for $k_{1z}$ and $k_{2z}$. The branch-cuts of $k_{1z}$ and $k_{2z}$ determine their correct signs and, depending on the value of $k_x$, it could happen that $k_{1z} + k_{2z} \neq 0$ at the points $\omega$ given by Eq.(31), in which case the two poles cease to exist.

---

Suppose we keep $k_x$ on the real axis $k_x'$ of the $k_x$-plane and choose the temporal frequency domain to be $\omega = \omega' + i\Omega$ (with $\Omega > 0$ to be subsequently determined). Then, in accordance with Eq.(28), when $k_x$ rises from zero to infinity along the $k_x'$-axis, the branch-points of $k_{1z}$ and $k_{2z}$ follow the trajectories depicted in Figs.(8a) and (8b), respectively. Here, each $k_z$ has two poles that are independent of $k_x$ and, therefore, remain fixed in their initial locations in Fig.8. In addition, each $k_z$ has four zeros, whose trajectories start either at the zeros of $\varepsilon(\omega)$ or at $\omega = 0$. All in all, the branch-points of $k_{1z}$ are in the lower half of the $\omega$-plane, whereas $k_{2z}$ has two branch-points in the upper half-plane. If we now pick $\Omega$ such that the straight-line $\omega = \omega' + i\Omega$ stays above all these branch-point trajectories, the response of the system to an incident wave-packet will turn out to be causal.

Whereas the spatial Fourier transform is evaluated over the real $k_x'$-axis, the temporal Fourier-Laplace transform must be evaluated on the straight line $\omega = \omega' + i\Omega$ (with $\Omega$ being large enough to place this straight line above all the branch-point trajectories depicted in Fig.8). For any given real-valued $k_x$, we may express $k_{1z}$ and $k_{2z}$ in the following way:

$$k_z = \sqrt{(\omega - \omega_1)(\omega + \omega_1^*)(\omega - \omega_2)(\omega + \omega_2^*)/(\omega - \omega_3)(\omega + \omega_3^*)}. \tag{32}$$

Defining the branch-cuts such that the argument of each of the six terms under the radical in Eq.(32) is in the interval $(-\pi, \pi]$, we find that $k_z \to \omega/c$ on a semi-circle of large radius $R$ in the upper-half $\omega$-plane. This is the all-important requirement for ensuring the causality of the signal arriving at points $(x, z)$ located further than $ct$ away from the interface (i.e., the $xy$-plane at $z = 0$). The reflected and transmitted waves can be computed by a two-dimensional Fourier transform, first along the real-axis $k_x'$, and then along the straight line $\omega = \omega' + i\Omega$ in the $\omega$-plane. The presence of the $k_{2z}$ branch-cuts in between this straight line and the real $\omega'$-axis is the reason why the temporal Fourier transform cannot be carried out directly along the real $\omega'$-axis.

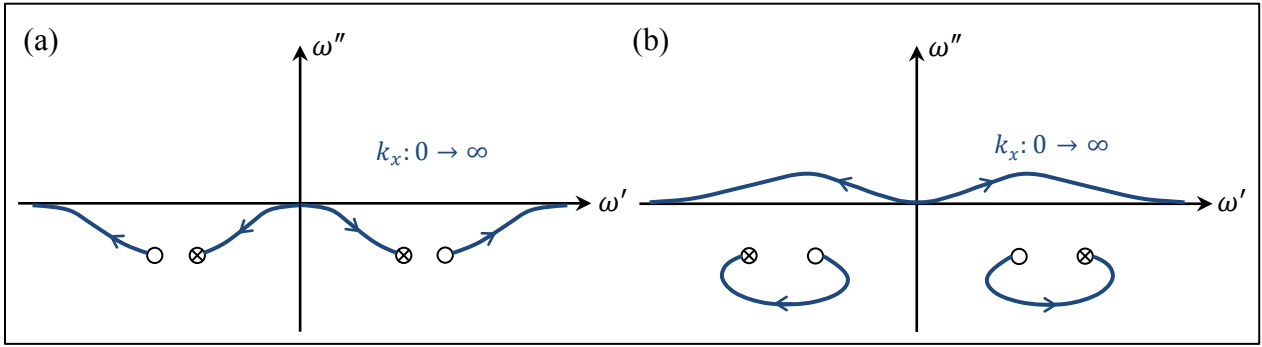

**Fig.8**. Trajectories of the branch-points for (a) $k_{1z}$ and (b) $k_{2z}$ in the $\omega$-plane, when the real-valued $k_x$ rises from zero to infinity. Each $k_z$ has two poles and four zeros, which appear as pairs of mirror-images in the $\omega''$-axis. The poles of $k_{1z}$ and $k_{2z}$ are fixed in place, whereas their zeros are functions of $k_x$.

An alternative method of computing the reflected and/or transmitted waves involves a deformation of the integration contour in the $k_x$-plane, which would make it possible to evaluate the temporal Fourier integral along the real $\omega'$-axis. As shown in Fig.9, the deformed integration contour is identified by plotting the $k_x$-plane trajectories of all the zeros of $k_{1z}$ and $k_{2z}$, when $\omega'$ is fixed (at a positive value) while $\omega''$ rises from zero to infinity along a straight line parallel to the imaginary axis of the $\omega$-plane. The $k_x$-plane integration contour is subsequently chosen such that it avoids crossing any and all such trajectories of the branch-points of $k_{1z}$ and $k_{2z}$ within the $k_x$-plane.

Computing the Fourier integral over the deformed contour in the $k_x$-plane requires a choice for the branch-cuts. In general, $k_z$ can be written as follows:

$$k_z = \pm i \bigl[k_x + (\omega/c)\sqrt{\varepsilon(\omega)}\,\bigr]^{\frac{1}{2}} \bigl[k_x - (\omega/c)\sqrt{\varepsilon(\omega)}\,\bigr]^{\frac{1}{2}}. \tag{33}$$

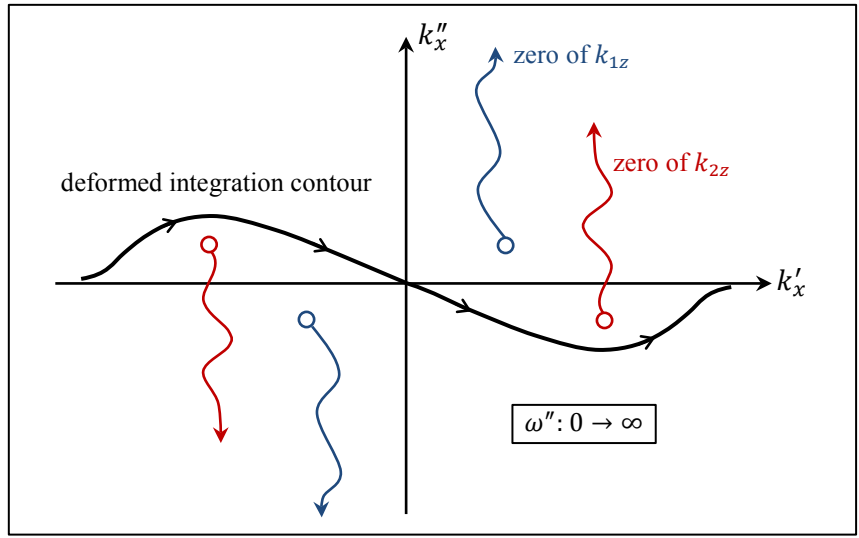

**Fig.9**. Trajectories of the zeros of $k_{1z}$ and $k_{2z}$ in the $k_x$-plane, when $\omega'$ is fixed at a positive value while $\omega''$ rises from zero to infinity. The deformed integration contour is chosen such that it avoids crossing any and all of these trajectories for all positive values of $\omega'$.

For a fixed, positive, real-valued $\omega$, the branch-cuts associated with $k_{1z}$ and $k_{2z}$ are shown in Fig.10. The solid red arrows in this figure represent $k_x \pm (\omega/c)\sqrt{\varepsilon_2(\omega)}$, whose phase angles $\varphi$ are measured in conjunction with their corresponding branch-cut. The correct choice of sign for $k_z$ of Eq.(33) must be made in the context of causality to ensure that $k_z \to \omega/c$ when $\omega'' \to \infty$ in the upper-half $\omega$-plane. Considering the range of values of $\varphi$ shown in Fig.10, the correct sign for $k_z$ of Eq.(33) is the *minus* sign.

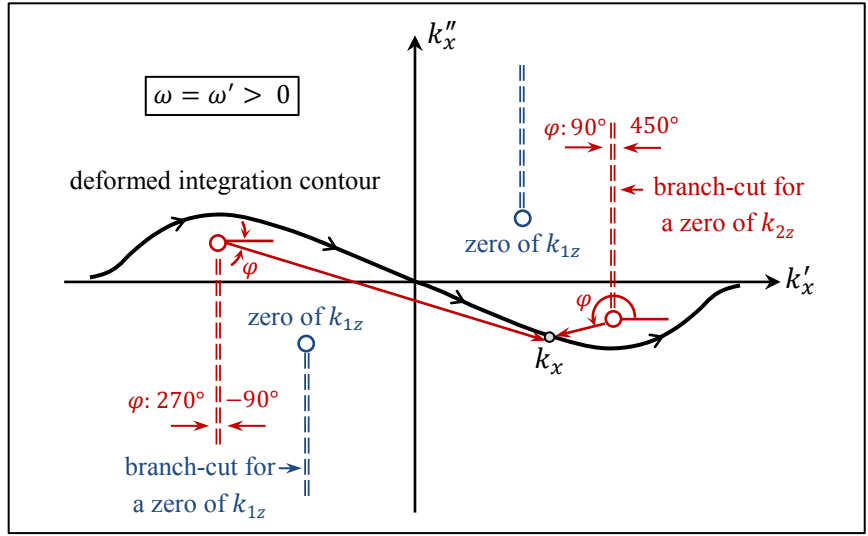

**Fig.10**. Branch-cuts for $k_{1z}$ and $k_{2z}$ in the $k_x$-plane corresponding to a fixed, positive, real-valued frequency $\omega$. The branch-cuts are chosen to ensure that they do not cross the deformed integration path. For a specific (real-valued) $\omega$ and a specific point $k_x$ on the integration contour, the solid red arrows represent the two terms whose square root appears in Eq.(33) in the expression of $k_{2z}$.

**7. Issues associated with gainy slab calculations**. We are now in a position to describe the general problem of reflection and transmission for gainy slabs of finite thickness $d$ depicted in Fig.1(b). As before, the finite duration of the incident light pulse demands that we account for the frequency-dependence of the dielectric functions of the three media (i.e., incident medium 1, gain medium 2, and transmittance medium 3). This we do by relying upon the single-oscillator Lorentz model of the three dielectric functions, as follows:[7,8]

$$\varepsilon_n(\omega) = 1 \pm \frac{\omega_{pn}^2}{\omega_{rn}^2 - \omega^2 - i\gamma_n \omega}. \tag{34}$$

Each medium has its own plasma frequency $\omega_{pn}$, resonance frequency $\omega_{rn}$, and damping coefficient $\gamma_n$. These are all real-valued and positive constants, although, on physical grounds, we expect $\gamma_n \ll \omega_{rn}$. The plus sign in Eq.(34) applies to the incidence and transmittance media ($n = 1, 3$), which are passive, whereas the minus sign applies to the gain medium ($n = 2$), which is active. In the case of the passive media, both poles and both zeros of $\varepsilon_n(\omega)$ are in the lower-half of the $\omega$-plane, located on a straight line slightly below the real-axis. In the case of the gain medium, the poles of $\varepsilon_2(\omega)$ are in the lower-half plane, but one of its zeros may creep up into the upper-half plane if one chooses $\omega_{p2} \geq \omega_{r2}$. (In our numerical work[1] we used the same parameter values for media 1 and 2 as already mentioned in Sec.6, and for medium 3 we set $\omega_{p3} = 4$, $\omega_{r3} = 16$, $\gamma_3 = 1$.)

To keep track of the symmetries of the various functions appearing in these calculations, one should remember that, in general, $\omega = \omega' + i\Omega$ (with $\Omega \geq 0$) and that, in accordance with Eq.(34), the media 1, 2, and 3 satisfy the identity $\varepsilon_n(-\omega^*) = \varepsilon_n^*(\omega)$. It is also essential to remember that $\omega/c$ must be taken out of the square root expressions pertaining to $k_{1z}$, $k_{2z}$, and $k_{3z}$, so that one could always write

$$k_z(k_x, \omega) = (\omega/c)\sqrt{\varepsilon(\omega) - (ck_x/\omega)^2} \quad \rightarrow \quad k_z(-k_x^*, -\omega^*) = -k_z^*(k_x, \omega). \tag{35}$$

For an $s$-polarized incident plane-wave, the Fresnel reflection coefficient at the front facet of the slab (i.e., at $z = 0$) and the transmission coefficient at its rear facet (i.e., at $z = d$) are given by[2]

$$\rho(k_x, \omega) = E_y^{(\text{ref})} / E_y^{(\text{inc})} = \frac{\rho_{12} + \rho_{23} \exp(2ik_{2z}d)}{1 - \nu}, \tag{36}$$

$$\tau(k_x, \omega) = E_y^{(\text{trans})} / E_y^{(\text{inc})} = \frac{(1 + \rho_{12})(1 + \rho_{23}) \exp(ik_{2z}d)}{1 - \nu}. \tag{37}$$

Here, the roundtrip coefficient $\nu$, itself a function of $k_x$ and $\omega$, is

$$\nu = \rho_{21}\rho_{23} \exp(2ik_{2z}d), \tag{38}$$

while the Fresnel reflection coefficients at the various interfaces are given by

$$\rho_{12} = -\rho_{21} = (k_{1z} - k_{2z})/(k_{1z} + k_{2z}), \tag{39}$$

$$\rho_{23} = (k_{2z} - k_{3z})/(k_{2z} + k_{3z}). \tag{40}$$

Note that Eq.(35) now guarantees the important symmetry relations $\rho(-k_x^*, -\omega^*) = \rho^*(k_x, \omega)$ and $\tau(-k_x^*, -\omega^*) = \tau^*(k_x, \omega)$. (If need be, an expression for $E$-field inside the slab at $0 \le z \le d$ may similarly be written as well.[1])

The goal is to compute the $E$-field at arbitrary points $(x, z, t)$ in spacetime. Considering that the incident light pulse (or wavepacket) has finite duration in time and finite width along the $x$-axis, the incident $E$-field must be written as a superposition of plane-waves having all (real-valued) frequencies $\omega$ and all (real-valued) $x$-directed components $k_x$ of the incident $k$-vector; that is,

$$E_y^{(\text{inc})}(x, z = 0, t) = (2\pi)^{-2} \iint_{-\infty}^{\infty} \mathcal{E}(k_x, \omega) \exp[i(k_x x - \omega t)] \, dk_x d\omega. \tag{41}$$

In general, the Fourier transform of a finite-width, square-integrable function is known to be analytic everywhere in the corresponding complex-frequency plane. Therefore, we may treat the incident spectrum $\mathcal{E}(k_x, \omega)$ as an analytic function of $\omega = \omega' + i\omega''$. Similarly, we may consider $\mathcal{E}(k_x, \omega)$ to be an analytic function of $k_x = k_x' + ik_x''$. Our crucial first step is to work with frequencies $\omega = \omega' + i\Omega$, where $\Omega > 0$ defines a sufficiently large (but fixed) value for the imaginary part $\omega''$ of our complex frequencies. (This is perhaps the most important step in the entire process, which was originally suggested by Briggs[10] and was strongly advocated, decades later, by Skaar and his collaborators[11-13]).

At this stage we have not yet moved into the complex $k_x$-plane, so the integral in Eq.(41) is over the entire real axis $k_x'$ and the entire straight line $\omega = \omega' + i\Omega$, which is parallel to the real axis $\omega'$ at a fixed height $\Omega$ above the real axis of the $\omega$-plane. Since the $E$-field distributions of interest are always real-valued, it turns out that we do not need to worry about the negative values of $\omega'$, as everything on the left-half of the straight line located at $\omega'' = \Omega$ turns out to be the complex-conjugate of the corresponding entity on the right-half of the same line. All we need to do then is to integrate the functions of interest over the right-half of this straight-line in the $\omega$-plane, then take the real part of the resulting integral. It is important to remember, however, that the integral over $k_x'$ is computed over the entire $k_x'$-axis, that is, from $-\infty$ to $\infty$. Consequently, in what follows, the range of values of $\omega'$ is $[0, \infty)$ while the range of values of $k_x'$ is $(-\infty, \infty)$, with the caveat that, in the end, it is the real part of the overall integral (taken first over $k'_x$, then over $\omega'$) that needs to be evaluated.

Next, we deform the integration path in the $k_x$-plane so that, instead of being evaluated over the real-axis $k'_x$, the $k_x$ integral will be computed over a contour such as $C$ shown in Fig.11(b). Since $\mathcal{E}(\omega, k_x)$ is analytic throughout the entire $k_x$-plane, this change of the integration contour will not, according to Cauchy's theorem,[4-6] affect the final result. At his point, the integral in Eq.(41) can be evaluated on any combination of paths, that is, one in the $\omega$-plane (say, the real axis $\omega'$, or the parallel straight line located at $\omega'' = \Omega$), and another one in the $k_x$-plane (say, the real axis $k'_x$, or the arbitrarily deformed contour $C$).

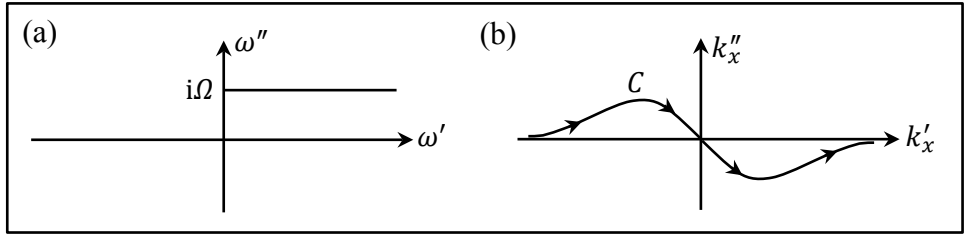

**Fig.11**. Alternative contours of integration for the incident wavepacket described by the inverse Fourier integral in Eq.(41). (a) Switching from Fourier to Laplace transform requires a shift of the integration path in the $\omega$-plane from the real axis $\omega'$ to a parallel line located at $\omega'' = \Omega$. Since the $E$-field distributions of interest are real-valued, one needs to evaluate the integral only over the right-half of this straight line, i.e., from $\omega = 0 + i\Omega$ to $\omega = \infty + i\Omega$, provided that, in the end, the desired result is taken to be the real part of the integral. (b) By the same token, since $\mathcal{E}(\omega, k_x)$ is an analytic function of $k_x$ in the entire $k_x$-plane, one is justified in changing the integration path from the real-axis $k'_x$ to an arbitrary contour such as $C$.

We postpone a description of the selection process for $C$ until after the following discussion of how to incorporate a response function (such as the Fresnel reflection coefficient of the slab) into the integrand of Eq.(41). Suffice it to say at this point, that the choice of $C$ will be dictated by the desire to eliminate from the strip $0 \leq \omega'' \leq \Omega$ of the $\omega$-plane all the branch-cuts (if any) as well as all the singularities of the response function (or at least as many singularities as possible). Assuming that such a contour $C$ exists (and that it can be identified), one can return the integration path in the $\omega$-plane back to the real axis $\omega'$, then compute the $E$-field distribution at the desired location $(x, z, t)$ by integrating the corresponding integrand, first over the contour $C$ in the $k_x$-plane (with the integrand being evaluated at a fixed, positive, real $\omega = \omega'$), and then over the positive half of the real axis in the $\omega$-plane, that is, from $\omega' = 0$ to $\infty$ (at $\omega'' = 0$).

In the next step, we multiply the incident spectrum $\mathcal{E}(k_x, \omega)$ with the response function of interest, say, the Fresnel reflection coefficient $\rho$ given by Eq.(36). In general, such functions depend on both $k_x$ and $\omega$, so, from now on, we shall assume that the integrand of Eq.(41) is multiplied by $\rho(k_x, \omega)$, keeping in mind that the Fresnel coefficients, being exact solutions of Maxwell's equations, are valid for complex $\omega$ and, simultaneously, for complex $k_x$. Needless to say, the multiplication of the integrand of Eq.(41) by $\rho(k_x, \omega)$ is exemplary and, in principle as well as in practice, the $E$-field at any point $(x, z)$ inside or outside the slab can be computed by precisely the same methods as described here.

We are finally in a position to describe the general numerical procedure for constructing the contour $C$. Start on the $\omega'$-axis in the $\omega$-plane and move straight up (i.e., parallel to the $\omega''$-axis). At each point $\omega$ on this straight, vertical half-line, find all (complex) values of $k_x$ that are either a branch point or a singularity (i.e., a pole) of the integrand. If, at a given point $\omega$, several such values of $k_x$ are found, each would belong to a different trajectory in the $k_x$-plane. As you move up in the $\omega$-plane, you must plot (in the $k_x$-plane) all the $k_x$ trajectories thus found. Note that, for any such trajectory on the right-half of the $k_x$-plane, there will be a corresponding trajectory on the left-half, since the response functions contain only $k_x^2$, which indicates that both $\pm k_x$ are solutions of the relevant equations. For a given value of $\omega'$ on the real axis of the $\omega$-plane, a trajectory typically starts somewhere above or below the $k_x'$-axis in the $k_x$-plane, then moves upward (on the right-hand side of the $k_x$-plane) or downward (on the left-hand side). In general, you will have one or more trajectories that start somewhere below the positive $k_x'$-axis and move upward, whereas their flipped image(s) will start above the negative $k_x'$-axis and move downward. You should construct all these $k_x$-plane trajectories for all the vertical half-lines within the $\omega$-plane that originate at various points $\omega' > 0$ on the real-axis.

Once the $k_x$-plane is populated with all the aforementioned trajectories, construct the contour $C$ such that it would stay below the starting points of all the trajectories in $Q_4$ of the $k_x$-plane, and above the corresponding starting points in $Q_2$. The expected shape of the contour $C$, as shown in Fig.11(b), is symmetric relative to the origin of the $k_x$-plane.

So long as the integration path in the $k_x$-plane stays on the contour $C$ thus constructed, it will avoid all the branch-cuts and all the singularities within the strip $0 \le \omega'' \le \Omega$ of the $\omega$-plane. This is why, in light of the Cauchy theorem, the integration path in the $\omega$-plane can be returned to the real-axis $\omega'$, provided, of course, that the contour $C$ is the properly chosen path of integration in the $k_x$-plane.

In the following paragraphs, we elaborate some details of the computation method just described. These paragraphs should give the reader a better grasp of the ideas involved, or may help to simplify the implementation of the algorithm under certain circumstances.

**7.1. Deforming the integration path in the complex $k_x$-plane**. The Fourier transform $F(k_x)$ of the finite-width, square-integrable function $f(x)$ is analytic in the entire complex plane $k_x = k_x' + \mathrm{i} k_x''$. As such, the integration path for the inverse Fourier integral can be deformed away from the real axis $k_x'$ into any (arbitrary) contour $C$ that forms a closed loop with the $k_x'$-axis, as follows:

$$f(x) = (2\pi)^{-1} \int_{-\infty}^{\infty} F(k_x') \exp(\mathrm{i} k_x' x) \, \mathrm{d} k_x' = (2\pi)^{-1} \int_C F(k_x) \exp(\mathrm{i} k_x x) \, \mathrm{d} k_x. \qquad (42)$$

**7.2. Reflected wave-packet**. The reflected beam at the front facet of the slab (i.e., at $z = 0$) is readily found to be

$$E_y^{(\mathrm{ref})}(x,t) = 2(2\pi)^{-2} \, \mathrm{Re}\left[ \int_{\omega=0}^{\infty} \mathrm{d}\omega \, e^{-\mathrm{i}\omega t} \int_{\text{entire } k_x \text{ contour}} \rho(k_x,\omega) \mathcal{E}(k_x,\omega) e^{\mathrm{i} k_x x} \mathrm{d} k_x \right]. \qquad (43)$$

Note that the $k_x$-integral is taken over both halves of the contour located in $Q_2$ and $Q_4$ of the $k_x$-plane, whereas the integral over $\omega$ is an ordinary inverse Fourier transformation taken along the real $\omega$-axis. The symmetry of the problem allows the $\omega$-integral to be taken over the positive values of $\omega$ only, as indicated in Eq.(43). However, with reference to Eq.(25), the long tail of $\tilde{f}(k_x - k_{xc})$ as well as that of $\tilde{g}(\omega + \omega_c)$ require explicit inclusion in the $k_x$-integral of that part of the integration contour that resides in $Q_2$ of the $k_x$-plane.

**7.3. Branch-cuts play no role in constructing the contour $C$ for a finite-thickness slab**. For a gainy slab of finite thickness $d$, there would be no need to deform the integration contour in the $k_x$-plane if the sole purpose were to eliminate the branch-cuts from the upper-half $\omega$-plane. This is because, for $k_{2z}$ of the gain medium, both $\pm k_{2z}$ yield identical expressions for the Fresnel reflection and transmission coefficients.[1,14] In other words, even though branch-cuts are generally needed for a unique specification of $k_{2z}$, the Fresnel reflection and transmission coefficients (as well as the field amplitudes inside the slab) would be analytic everywhere within the complex $k_x$- and $\omega$-planes were it not for the singular points (i.e., poles) of the relevant integrand. It may be argued that one must still be careful with the $k_z$ values inside the incidence and transmittance media, since correct signs for $k_{1z}$ and $k_{3z}$ are needed in the calculations. However, given that both the incidence and transmittance media are passive, so long as the integration contour $C$ is properly chosen to avoid the trajectories of the poles of the integrand within the $k_x$-plane, it can be shown that no branch-cuts for $k_{1z}$ and $k_{3z}$ will creep into the upper-half $\omega$-plane.

**7.4. Constructing the integration contour $C$ in the $k_x$-plane**. With reference to Fig.12, for each value of $\omega'$ on the positive real-axis of the $\omega$-plane, start at $\omega'' = 0$, then move up parallel to the imaginary axis $\omega''$ and identify the trajectory of points in the $k_x$-plane that correspond to a singularity of the integrand. Once the entire first quadrant of the $\omega$-plane is surveyed in this way, the corresponding domain of singularities in the $k_x$-plane will be identified. Note that each singularity is associated with both $\pm k_x$. Subsequently, the contour $C$ is constructed such that it bypasses all the singularities. In general, $C$ should go through the origin of the $k_x$-plane and be anti-symmetric relative to this point, lying entirely in $Q_2$ and $Q_4$ of the $k_x$-plane. Repeating the procedure for the negative values of $\omega'$ yields the complex conjugate of the contour $C$, which lies in $Q_1$ and $Q_3$ of the $k_x$-plane.

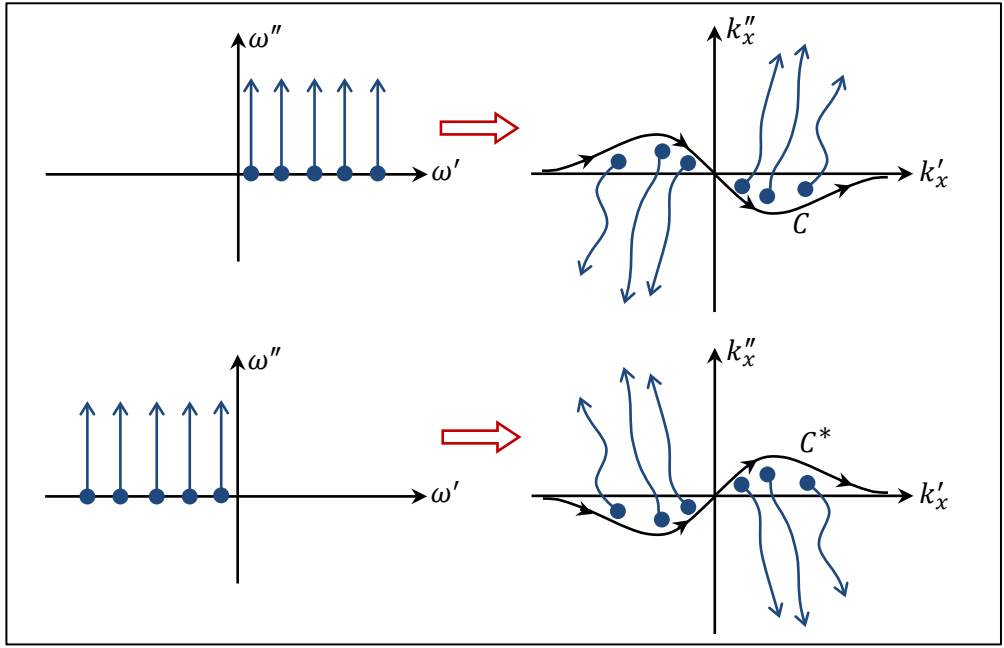

**Fig. 12**. Method of constructing the integration contour $C$ in the $k_x$-plane. Note that, for each point $\omega = \omega' + i\omega''$, whether in $Q_1$ or $Q_2$ of the $\omega$-plane, both $\pm k_x$ represent the same singularity. Note also that, for $\omega$ in $Q_2$, the singularities of $k_x$ are the complex-conjugates of those for the $Q_1$ values of $\omega$.

In practice, there will be no need to account for $Q_2$ of the $\omega$-plane, since, for every positive $\omega'$ and for fixed values of the $x$ and $z$ coordinates, the integral with respect to $k_x$ will be taken over the entire contour $C$. The inverse Fourier integral for a given $t$ is then evaluated over the positive $\omega'$-axis, with the real part of the integral yielding the desired signal at the spacetime location $(x, z, t)$.

**8. Concluding remarks**. This paper is a companion to another paper[1] that we have recently published on the subject of Fresnel reflection and transmission at the interface between a passive (i.e., transparent or partially absorbing) and an active (i.e., amplifying or gainy) material medium. Some of the theoretical background as well as physical and logical reasoning that went along with the numerical results reported in [1] were lacking in detail, hence the need for this companion paper. We have now covered the necessity of resorting to the Fourier-Laplace transformation for solving the particular problem under consideration, and also described some of the basic properties of this mathematical operation. The guiding principle behind the selection rule for the sign of $k_z$ (i.e., the $z$-component of the propagation vector $\boldsymbol{k}$ associated with each and every plane-wave involved in the analysis) has been the necessity of ensuring that the causality constraint[2,3,12] is satisfied. This has required that we choose for the incident beam a starting point in time, say, $t = t_0$, at which it arrives at the interfacial $xy$-plane located at $z = 0$, then proceed to exploit the relativistic principle that no information-carrying signal can reach the point $(x, y, z)$ prior to $t = t_0 + |z|/c$, where $c$ is the speed of light in vacuum. Enforcement of the causality constraint has meant that we must ensure the analyticity of the functions that appear as the integrand of the inverse Fourier-Laplace transform integrals in the upper parts of the complex $\omega$-plane. This, in turn, has led to a detailed investigation of the branch-points and branch-cuts of $k_z$ in the complex $\omega$-plane, and also in the complex $k_x$-plane, in search of appropriate (deformed or displaced) integration contours that avoid crossing paths with these branch-cuts.

Our original motivation for studying the Fresnel reflection and transmission at the interface between a passive and an active dielectric medium was to answer a question posed by the late Anthony Siegman in the context of total internal reflection (TIR) at such junctions.[15] It is a well-known fact that an electromagnetic plane-wave arriving at oblique incidence at the flat interface between a high-index dielectric host (the incidence medium) and a second, lower-index dielectric (the transmittance medium) will undergo TIR at sufficiently large angles of incidence. Under such circumstances, the electromagnetic field inside the low-index dielectric is taken to be an evanescent plane-wave (when this dielectric medium is transparent) or an "evanescent-like" plane-wave (when this dielectric medium is partially absorptive). In both cases, the electromagnetic wave residing inside the transmittance medium is assumed to decay exponentially away from the interface, lest it violates the principle of conservation of energy. The difficulty with allowing gain within the transmittance medium is that it will no longer support the argument based on energy conservation, hence the need for invoking the other powerful and universally-applicable principle of causality. (See [14] for a conventional treatment of TIR in the presence of gain and its associated difficulties.)

Our treatment of the Fresnel reflection and transmission at the interface between a transparent dielectric and a gain medium has now gone beyond the original question of TIR at such an interface to include cases involving finite-duration and finite-footprint wavepackets arriving at arbitrary angles of incidence. Also, in addition to the single-surface configuration, where the gain medium is semi-infinite, we have examined the case of a finite-thickness gainy slab sandwiched between two semi-infinite passive media. The finite-thickness slab introduces additional singularities (in the form of poles, as opposed to branch-cuts) in the upper-half $\omega$-plane, which we have described here and, in greater detail, in [1]. The beauty of the employed methods (involving the Fourier-Laplace transformation, deformed integration contours, the causality constraint, etc.) is that they can handle a variety of situations within a unified theoretical framework amenable to numerical computations.

**Acknowledgement**. The authors acknowledge valuable discussions extending over many years with Tobias Mansuripur, Johannes Skaar, Ewan Wright, and the late Anthony Siegman.

# Appendix A
# Branch-cuts of certain complex functions

A branch-cut in the complex $z$-plane, where $z = |z|\exp(i\varphi)$, limits the range of the phase-angle $\varphi$ to a $2\pi$ interval, so that certain functions defined over the entire complex-plane (or a subset thereof) will end up having unique values. For instance, the natural logarithm function

$$\ln z = \ln\{|z|\exp[i(2m\pi + \varphi)]\} = \ln|z| + i(2m\pi + \varphi), \tag{A1}$$

needs a branch-cut such as that over the negative real-axis (see Fig.A1) that limits the range of the imaginary part of $\ln z$ to the interval $(-\pi, \pi]$. In fact, any line, straight or meandering, starting at the origin and heading to infinity along a more-or-less arbitrary path, can be used to define a branch-cut for $\ln z$, ensuring that each $z$ is assigned a unique natural logarithm depending on the crossing point of the branch-cut with a circle of radius $|z|$ centered at the origin. In general, a function defined with the aid of a branch-cut cannot be analytic on the branch-cut, because for any $z$ located on the branch-cut, any small neighborhood contains points on both sides of the cut.

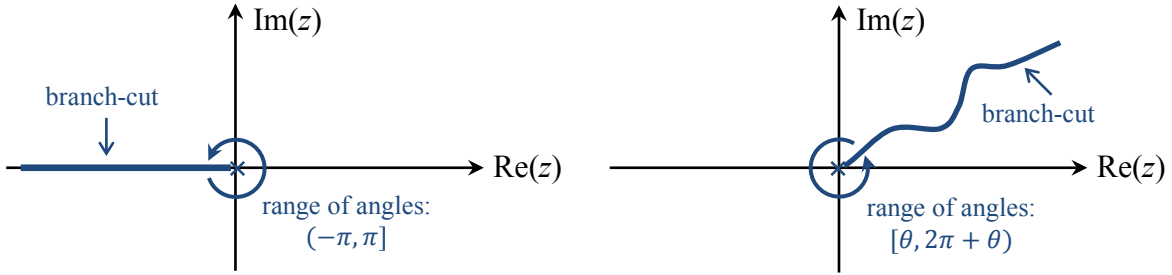

**Fig.A1**. A branch-cut is a straight or meandering line starting from a point of singularity and heading to infinity along a more-or-less arbitrary path in the complex plane. The phase $\varphi$ of the complex number $z = |z|\exp(i\varphi)$ is thus confined to a $2\pi$ interval. Note that the meandering branch-cut must cross a circle of arbitrary radius centered at the origin at one and only one point.

Here, we investigate the branch-cut(s) of the function $f(z) = P_n^\alpha(z)$, where $P_n(z) = \sum_{m=0}^n a_m z^m$ is an $n^{\text{th}}$-order polynomial having arbitrary complex coefficients, while $\alpha$ is a real-valued, non-integer constant. Since, in general, any $n^{\text{th}}$-order polynomial has $n$ zeros in the complex-plane, one is allowed to write it as $P_n(z) = a_n \prod_{m=1}^n (z - z_m)$, that is, as the product of $n$ terms such as $(z - z_1)$, $(z - z_2)$, etc. We limit the following discussion to a few simple polynomials such as $z^\alpha$, $(z - z_\text{o})^\alpha$, and $[(z - z_1)(z - z_2)]^\alpha$, which should suffice to illustrate the application of the concepts and ideas to more complicated cases as well.

The function $z^\alpha$, where $\alpha$ is a real, non-integer constant, needs a branch-cut if it is to be uniquely-defined for each $z$. One way to see this is to recognize that $z^\alpha = \exp(\alpha \ln z)$, and that $\ln z$ requires a branch-cut. (Of course, if $\alpha$ happens to be an integer $n$, then $z^n = |z|^n \exp(in\varphi)$ is well-defined and there will be no need to resort to the logarithmic function.) Alternatively, we may write $z^\alpha = |z|^\alpha \exp[i\alpha(2m\pi + \varphi)]$, and note that different values of $m$ could give rise to different phase-angles $2\alpha m\pi$ when $\alpha$ is *not* an integer. Any straight or meandering line drawn in the complex-plane, starting at the origin and heading to infinity along a path that continually recedes from the origin, would provide an appropriate branch-cut for $f(z) = z^\alpha$.

The function $f(z) = \sqrt{z - z_\text{o}}$, where $z_\text{o}$ is a fixed point in the complex-plane, may be specified as follows:

i) Draw a straight-line connecting $z_\text{o}$ to $z$ in the complex-plane.

ii) Determine the magnitude $|z - z_0|$ and phase-angle $\varphi$ of $z - z_0$, where $0 \leq \varphi < 2\pi$.

iii) Write $\sqrt{z - z_0} = |z - z_0|^{\frac{1}{2}} \exp[i(m\pi + \frac{1}{2}\varphi)]$.

It is clear that, depending on the integer $m$ being odd or even, the above square-root will have two different values. A branch-cut in the form of a straight or meandering line, starting from $z_0$ and heading to infinity along a path that continually recedes from $z_0$, is thus needed to uniquely specify the square root of $z - z_0$. The phase of $z - z_0$ is no longer taken to be in the interval $[0, 2\pi)$; instead, for each $z$, it must be chosen within a different $2\pi$-interval referenced to the crossing-point of the (meandering) branch-cut with a circle of radius $|z - z_0|$ centered at $z_0$.

The function $\sqrt{(z - z_1)(z - z_2)}$, where $z_1$ and $z_2$ are two fixed points in the complex-plane, can be uniquely specified with the aid of two branch-cuts, one starting at $z_1$, the other at $z_2$, both moving along different (arbitrary) paths to infinity. The two paths, of course, may join at some point in the complex-plane, then overlap for the rest of their journey to infinity. While this overlapping section of the two branch-cuts is still needed for the unique identification of the phase angles of $\sqrt{z - z_1}$ and $\sqrt{z - z_2}$, it no longer serves as a boundary between two regions of analyticity. This is because the $\pi$ phase-jumps of $\sqrt{z - z_1}$ and $\sqrt{z - z_2}$ cancel out across the overlapping segment of the two branch-cuts. The boundary separating different regions of analyticity is thus confined to the (straight or meandering) line connecting $z_1$ to $z_2$. Note that the overlapping segments of the two branch-cuts would remain a boundary between two regions of analyticity if the function, instead of being the square-root, happens to be the cube-root (or some other root) of $(z - z_1)(z - z_2)$.

Similar considerations apply to functions of the form $f(z) = P_n^\alpha(z)/P_m^\beta(z)$, where $\alpha$ and $\beta$ are real-valued non-integers. The function now has $n$ zeros and $m$ poles, each of which could serve as the starting point of a straight or meandering branch-cut. As before, two or more branch-cuts are allowed to have overlapping segments. In general, these overlapping segments may or may not act as boundaries separating adjacent regions of analyticity, depending on whether the overall phase of $f(z)$ remains continuous or discontinuous across the overlapping part of two or more branch-cuts.

## Appendix B
### $\omega$-plane trajectories of the poles and zeros of $k_z$ when real-valued $k_x$ goes from 0 to ∞

The poles and zeros of the function $k_z = \sqrt{\mu(\omega)\varepsilon(\omega)(\omega/c)^2 - k_x^2}$ play important roles in determining the behavior of a homogeneous, linear, isotropic, lossy/gainy medium. Here we consider the case of non-magnetic media, having $\mu(\omega) = 1$, $\varepsilon(\omega) = 1 \pm \omega_p^2/(\omega_r^2 - \omega^2 - i\gamma\omega)$, with the plus (minus) sign corresponding to optical loss (gain). In general, the plasma frequency $\omega_p$ and the resonance frequency $\omega_r$ are real-valued and positive, and so is the loss/gain parameter $\gamma$. We may safely assume that $\omega_r \gg \gamma$ for cases of practical interest, and, in order to simplify the following analysis, we assume that $\omega_r^2 > (\omega_p^2 + \frac{1}{4}\gamma^2)$. We now write

$$k_z = \sqrt{\frac{(\omega-\omega_1)(\omega-\omega_2)(\omega/c)^2}{(\omega-\omega_3)(\omega-\omega_4)} - k_x^2}, \quad (B1)$$

where

$$\omega_{1,2} = -\tfrac{1}{2}i\gamma \pm \sqrt{\omega_r^2 \pm \omega_p^2 - \tfrac{1}{4}\gamma^2}, \qquad \omega_{3,4} = -\tfrac{1}{2}i\gamma \pm \sqrt{\omega_r^2 - \tfrac{1}{4}\gamma^2}. \quad (B2)$$

Clearly, $\omega_1, \omega_2, \omega_3$ and $\omega_4$ are in the lower-half $\omega$-plane, all aligned on a straight line parallel to the real-axis $\omega'$, as shown in Fig.B1. Note that $\omega_2 = -\omega_1^*$, $\omega_4 = -\omega_3^*$, and that $\omega_1' > \omega_3'$ for lossy media, whereas $\omega_1' < \omega_3'$ in the case of gainy media. The function $k_z(\omega)$ has two fixed poles at $\omega = \omega_{3,4}$, and four zeros whose locations vary in the $\omega$-plane depending on the (real-valued) parameter $k_x^2$; the trajectories of these zeros, as $|k_x|$ goes from 0 to $\infty$, are depicted in Fig.B1.

It is easy to see that the poles and zeros of $k_z$ appear in pairs, as mirror images of each other in the $\omega''$-axis. While all four zeros of $k_z$ in the case of lossy media are in the lower-half $\omega$-plane, in the case of gainy media, two of the zeros relocate to the upper half-plane. Unique identification of $k_z$ (for each value of $k_x$) thus requires the specification of six semi-infinite branch-cuts in the $\omega$-plane. Crossing each such branch-cut will cause $k_z$ to change sign (i.e., suffer a $\pi$ phase-shift). However, these branch-cuts may be paired together in order to eliminate their overlapping segments. One such pairing of the branch-cuts connects each pole or zero to its own mirror image in the $\omega''$-axis via a straight line drawn parallel to the $\omega'$-axis. Consequently, for a given value of $k_x$ (real), the function $k_z(\omega)$ will be analytic throughout the $\omega$-plane except on three straight-line segments parallel to the $\omega'$-axis. Needless to say, other pairings are also possible. For instance, one may draw two short, straight, line-segments connecting the poles $\omega = \omega_{3,4}$ to their nearest zeros in the lower half-plane; the remaining zeros will then be connected to each other by a third straight line-segment parallel to the $\omega'$-axis.

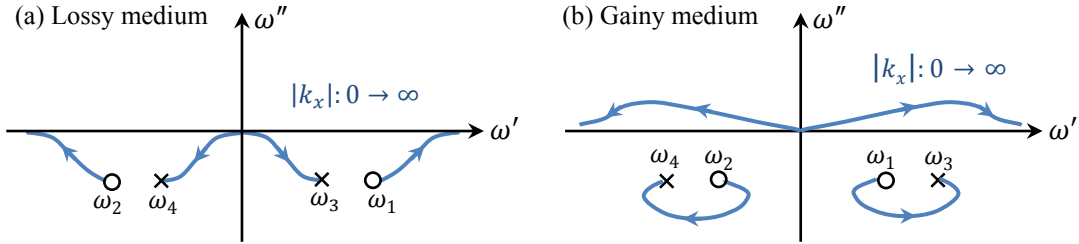

**Fig.B1**. Trajectories of the zeros of $k_z$ in the $\omega$-plane as the real-valued $k_x$ rises from 0 to $\infty$. (a) The case of a lossy medium, where $\omega_1' > \omega_3'$. (b) The case of a gainy medium, where $\omega_1' < \omega_3'$. Note that, at $k_x = 0$, the zeros of $k_z$ are at $\omega = 0$ (two overlapping zeros), $\omega = \omega_1$, and $\omega = \omega_2$. As $|k_x|$ rises from 0 to $\infty$, the zeros of $k_z$ move away from these points in such a way as to keep the phase angle of $(\omega - \omega_1)(\omega - \omega_2)(\omega/c)^2/[(\omega - \omega_3)(\omega - \omega_4)]$ equal to zero at all times.

## Appendix C
## Generalized Fourier-Laplace Transform

The Fourier transform of $f(x)$, a complex-valued function of the real-variable $x$, is usually defined as $F(k) = \int_{-\infty}^{\infty} f(x) \exp(-ikx)\, dx$, where $k$ is the real-valued Fourier variable. Here we allow $k$ to be a complex variable, and evaluate the inverse Fourier transform of $F(k)$ on a deformed integration path, namely, one that deviates from the real axis in the complex $k$-plane, as shown in Fig.C1. Assuming that the function $F(k)$ is analytic in the region that is enclosed by the deformed path and the real axis, we will have

$$f(x) = (2\pi)^{-1} \int_{\text{path}} F(k) \exp(+ikx)\, dk. \tag{C1}$$

That the above generalized inverse Fourier integral over the deformed path does indeed return the original function $f(x)$ may be confirmed by substituting for $F(k)$ in Eq.(C1) its definition as a Fourier integral, then changing the order of integration to arrive at

$$f(x) = (2\pi)^{-1} \int_{\text{path}} \left[ \int_{-\infty}^{\infty} f(x') \exp(-ikx')\, dx' \right] \exp(+ikx)\, dk \qquad \text{}$$

$$= (2\pi)^{-1} \int_{-\infty}^{\infty} f(x') \left\{ \int_{\text{path}} \exp[ik(x-x')]\, dk \right\} dx'. \tag{C2}$$

Since $\exp[ik(x-x')]$ is an analytic function of $k$, we may invoke Cauchy's theorem to replace the deformed segments along the integration path with the original path of the inverse Fourier integral along the real axis. The integral of $\exp[ik(x-x')]$ along the chosen path in the complex-plane thus becomes equal to its integral over $k$ along the real axis, which is equal to $2\pi\delta(x-x')$. Subsequently, the sifting property of the delta-function makes the right-hand-side of Eq.(C2) equal to its left-hand-side, thereby completing the proof.

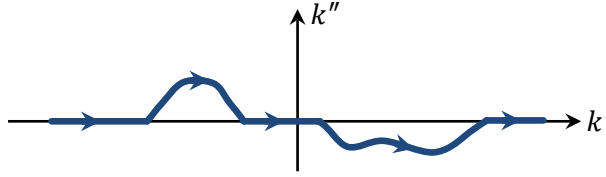

**Fig.C1**. The inverse Fourier transform of $F(k)$ is evaluated as an integral over a deformed path in the complex $k$-plane.

**Example 1**. The Fourier-Laplace transform of $f(x) = \text{rect}(x)$ is readily found to be

$$F(k) = \int_{-\frac{1}{2}}^{\frac{1}{2}} \exp(-ikx)\, dx = \frac{\exp(-\tfrac{1}{2}ik) - \exp(\tfrac{1}{2}ik)}{-ik} = \frac{\cosh(\tfrac{1}{2}k'')\sin(\tfrac{1}{2}k') + i\sinh(\tfrac{1}{2}k'')\cos(\tfrac{1}{2}k')}{\tfrac{1}{2}(k' + ik'')}. \tag{C3}$$

On the real axis, where $k'' = 0$, the above equation yields $F(k') = \text{sinc}(k'/2\pi)$, as expected. Note that the function $F(k)$ appears to have a single pole at $k = 0$. Upon closer inspection, however, we find that $\lim_{k \to 0} F(k) = \lim_{k',k'' \to 0} 2(\tfrac{1}{2}k' + \tfrac{1}{2}ik'')/(k' + ik'') = 1$. Consequently, $F(k)$ is an analytic function across the entire complex $k$-plane.

**Example 2**. The Fourier-Laplace transform of $f(x) = \exp[-(x - x_0)^2/w^2]$, where the real-valued constants $w > 0$ and $x_0$ are, respectively, the width and the center coordinate of the Gaussian function, is evaluated as follows:

$$F(k) = \int_{-\infty}^{\infty} \exp[-(x' - x_0)^2/w^2] \exp(-ikx')\, dx'$$

$$= \exp(-ikx_0)\exp(-\tfrac{1}{4}w^2k^2) \int_{-\infty}^{\infty} \exp[-(x' - x_0 + \tfrac{1}{2}i\, w^2 k)^2/w^2]\, dx'$$

$$= \exp(-ikx_0)\exp(-\tfrac{1}{4}w^2k^2) \int_{-\infty}^{\infty} \exp[-(x' - x_0 - \tfrac{1}{2}w^2 k'' + \tfrac{1}{2}i\, w^2 k')^2/w^2]\, dx'$$

$$= w\exp(-ikx_0)\exp(-\tfrac{1}{4}w^2k^2) \int_{-\infty}^{\infty} \exp[-(x + \tfrac{1}{2}i\, wk')^2]\, dx. \quad \boxed{\text{change of variable } x = (x' - x_0 - \tfrac{1}{2}w^2k'')/w} \tag{C4}$$

In the complex-plane $z = x + iy$, we now define a straight line $z = x + iy_0$ parallel to the $x$-axis, where $y_0 = \tfrac{1}{2}wk'$ is a constant. Application of Cauchy's theorem then yields

$$\int_{-\infty + iy_0}^{\infty + iy_0} \exp(-z^2)\, dz = \int_{-\infty}^{\infty} \exp(-x^2)\, dx = \sqrt{\pi}. \tag{C5}$$

The Fourier-Laplace transform of the Gaussian function $f(x)$ is thus seen to be $F(k) = \sqrt{\pi}w \exp(-ikx_0)\exp(-\tfrac{1}{4}w^2k^2)$, which is a well-defined function in the entire complex $k$-plane.

**General properties**. The Fourier-Laplace transform exhibits many of the well-known characteristics of the standard Fourier transform, as demonstrated by the following examples.

a) *Initial-value theorem*:  $\qquad f(0) = \int_{\text{path}} F(k) dk.$  (C6)

b) *Shift theorem*: $\qquad \int_{-\infty}^{\infty} f(x - x_0) \exp(-ikx) dx = \exp(-ikx_0) F(k).$  (C7)

c) *Scaling theorem*:  $\quad \int_{-\infty}^{\infty} f(\alpha x) \exp(-ikx) dx = |\alpha|^{-1} F(k/\alpha); \qquad (\alpha = \text{Real}).$  (C8)

d) *Differentiation theorem*: Here $f(x) \exp(k''x)$ must approach zero as $x \to \pm\infty$.

$$\int_{-\infty}^{\infty} f'(x) \exp(-ikx) dx = f(x) \exp(-ikx)|_{-\infty}^{\infty} + ik \int_{-\infty}^{\infty} f(x) \exp(-ikx) dx = ikF(k). \quad (C9)$$

e) *Convolution theorem*: $\qquad \int_{-\infty}^{\infty} \{\int_{-\infty}^{\infty} f(x')g(x-x')dx'\} \exp(-ikx) dx$

$$= \int_{-\infty}^{\infty} f(x')\{\int_{-\infty}^{\infty} g(x - x') \exp(-ikx) dx\} dx' = \int_{-\infty}^{\infty} f(x') \exp(-ikx') G(k) dx' = F(k)G(k).$$
(C10)

f) *Conjugation theorem*:

$$F^*(k) = \{\int_{-\infty}^{\infty} f(x) \exp(-ikx) dx\}^* = \int_{-\infty}^{\infty} f^*(x) \exp(ik^*x) dx. \quad (C11)$$

Note that $F^*(k)$ is equivalent to the Fourier-Laplace transform of $f^*(x)$ evaluated at $-k^*$. Applying the conjugation operation to the inverse transform integral yields

$$f^*(x) = (2\pi)^{-1}\{\int_{\text{path}} F(k) \exp(ikx) dk\}^* = (2\pi)^{-1} \int_{\text{path}} F^*(k) \exp(-ik^*x) dk^*. \quad (C12)$$

This is tantamount to flipping the integration path around the imaginary axis $k''$, which replaces $F^*(k)$ by the Fourier-Laplace transform of $f^*(x)$ evaluated at $-k^*$, replaces $\exp(-ik^*x)$ by $\exp(ikx)$, and replaces $dk^*$ by $-dk$. The end result is that the integral in Eq.(C12) yields the function $f^*(x)$, as it should.

**Laplace and Fourier transforms of causal time-domain functions**. Consider the function $f(t) = \text{step}(t) \cos(\omega_0 t)$. Its Laplace transform, valid for $\omega'' > 0$, is

$$F(\omega = \omega' + i\omega'') = \int_0^\infty \cos(\omega_0 t) \exp(i\omega t) dt = \frac{i}{2(\omega + \omega_0)} + \frac{i}{2(\omega - \omega_0)}. \quad (C13)$$

The Fourier transform of $f(t)$, however, cannot be obtained simply by setting $\omega'' = 0$; it requires a limit operation as follows:

$F(\omega = \omega') = \lim_{\alpha \to 0} \int_0^\infty \cos(\omega_0 t) \exp(-\alpha t) \exp(i\omega t) dt$

$$= \lim_{\alpha \to 0} \left[ \frac{½}{\alpha - i(\omega + \omega_0)} + \frac{½}{\alpha - i(\omega - \omega_0)} \right] = \lim_{\alpha \to 0} ½ \left[ \frac{\alpha + i(\omega + \omega_0)}{\alpha^2 + (\omega + \omega_0)^2} + \frac{\alpha + i(\omega - \omega_0)}{\alpha^2 + (\omega - \omega_0)^2} \right]$$

$$= ½\pi[\delta(\omega + \omega_0) + \delta(\omega - \omega_0)] + \frac{i}{2(\omega + \omega_0)} + \frac{i}{2(\omega - \omega_0)}. \quad (C14)$$

Figure C2(a) shows the integration contour for the inverse Laplace transform of $F(\omega)$ of Eq.(C13) in the case of $t > 0$. Each pole at $\omega = \pm\omega_0$ contributes to $\int_{-\infty+i\Omega}^{\infty+i\Omega} F(\omega)\exp(-i\omega t)\,d\omega$ a term equal to $\pi\exp(\pm i\omega_0 t)$, yielding $\cos(\omega_0 t)$ as the inverse Laplace transform for $t > 0$.

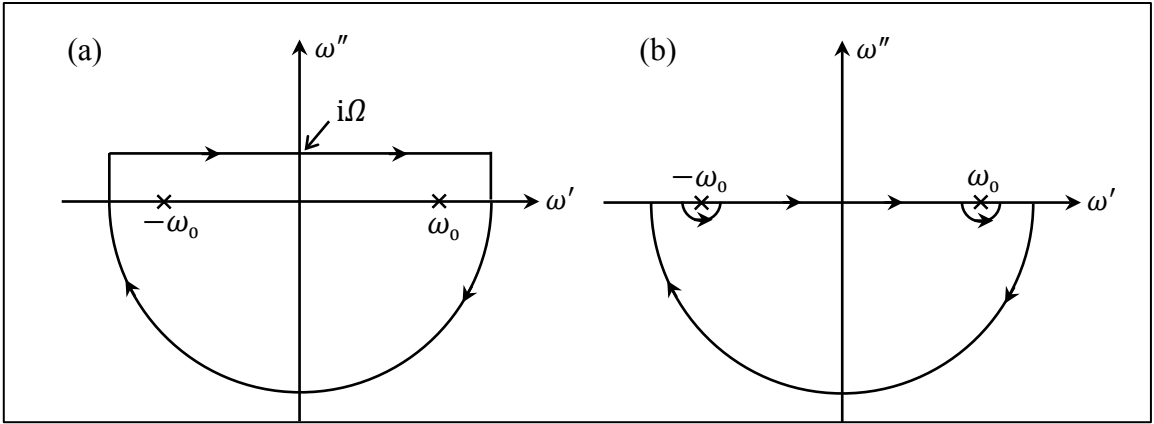

**Fig.C2**. Contours of integration for $t > 0$ for (a) the inverse Laplace transform of $F(\omega)$ given by Eq.(C13), and (b) the inverse Fourier transform of $F(\omega)$ given by Eq.(C14).

As for the inverse Fourier transform of $F(\omega)$ of Eq.(C14), the $\delta$-functions located at $\omega = \pm\omega_0$ contribute only $\tfrac{1}{2}\cos(\omega_0 t)$ to $(2\pi)^{-1}\int_{-\infty}^{\infty} F(\omega')\exp(-i\omega' t)\,d\omega'$. The contour depicted in Fig.C2(b) then shows that, for $t > 0$, the contribution of the remaining terms in Eq.(C14) is also $\tfrac{1}{2}\cos(\omega_0 t)$, resulting in the inverse Fourier transform being $\cos(\omega_0 t)$. For $t < 0$, the contour of Fig.C2(b) must be flipped around the horizontal axis, in which case the contribution of the small semi-circles will be $-\tfrac{1}{2}\cos(\omega_0 t)$, resulting in a net inverse Fourier transform of zero for $t < 0$.

As a second example consider the function $f(t) = \cos(\omega_0 t)$ in the time interval $[0, T]$; outside this interval the function vanishes, that is, $f(t) = 0$ when $t < 0$ and also when $t > T$. The Fourier and Laplace transforms of $f(t)$ may be readily calculated as follows:

$$F(\omega) = \int_0^T \cos(\omega_0 t)\exp(i\omega t)\,dt = \frac{\exp[i(\omega+\omega_0)T]-1}{2i(\omega+\omega_0)} + \frac{\exp[i(\omega-\omega_0)T]-1}{2i(\omega-\omega_0)}. \tag{C15}$$

In the case of the inverse Laplace transform, where $\omega'' = \Omega \neq 0$, each of the four terms appearing on the right-hand-side of Eq.(C15) requires a different contour of integration, depending on the sign of $\Omega$ and also whether $t < 0$, or $0 < t < T$, or $t > 0$. The large semi-circular contours will be either in the upper half-plane or, as in Fig.C2(a), in the lower half-plane.

In the case of the inverse Fourier transform, where $\omega'' = 0$, once again different terms in Eq.(C15) require different contours of integration for different values of $t$. For $t > T$, the contours will be the same as that in Fig.C2(b). All four terms in the inverse Fourier integral vanish over the large semi-circle in the lower half-plane. The contributions of the two small semi-circles will also be equal to zero because the residues of the exponential terms cancel out those of the constant terms appearing in the numerator of Eq.(C15). For $t < 0$, the contour of Fig.C2(b) must be flipped around the horizontal axis. Once again, the integrals over the large semi-circle in the upper half-plane will vanish, while the two integrals over each small semi-circle (also in the upper half-plane) will cancel out. For $0 < t < T$, the exponential terms will need a flipped contour, whereas the constant terms require the contour of Fig.C2(b). The contributions of the small semi-circles then add up to reproduce the function $f(t) = \cos(\omega_0 t)$.

# Appendix D
## A useful theorem of the Fourier transform theory

Functions that are limited in extent (or width) have an infinitely-wide Fourier transform, and vice-versa. The following theorem describes this particular characteristic of functions of finite width (or functions that have a finite bandwidth in the Fourier domain) in a more general setting.

*Theorem*: A bandlimited function $g(t)$, i.e., one whose Fourier transform $G(\omega)$ is square-integrable while confined to a finite interval $|\omega| \leq \omega_{max}$, cannot have a perfectly flat, contiguous region, i.e., an interval $(t_1, t_2)$ within which $g(t)$ equals a constant.

*Proof*: The inverse Fourier integral $g(t) = (2\pi)^{-1} \int_{-\omega_{max}}^{\omega_{max}} G(\omega) \exp(-i\omega t) \, d\omega$ can be extended to the entire complex-plane $t = t' + it''$, simply because, for any given value of $t''$, the function $G(\omega) \exp(\omega t'')$ is well-behaved over the finite interval $|\omega| \leq \omega_{max}$ and can, therefore, be Fourier transformed. What is more, $g(t)$ is differentiable everywhere in the complex $t$-plane, because $g'(t) = (2\pi i)^{-1} \int_{-\omega_{max}}^{\omega_{max}} \omega G(\omega) \exp(-i\omega t) \, d\omega$ should yield a finite value at any given $t$. Consequently, given that $g(t)$ is analytic in the entire $t$-plane, it must be expandable in a Taylor series around any arbitrary point $t$. However, if $g(t)$ happens to be flat over an interval $(t_1, t_2)$ on the real axis, it cannot have a Taylor series at any point $t$ within that interval. The proof that bandlimited functions cannot have flat regions is now complete.

# Appendix E
## Two-dimensional functions having a finite width in both dimensions

Consider the envelope functions $f(x)$ and $g(t)$, where $f(x) = 0$ outside the interval $(-\tfrac{1}{2}X, \tfrac{1}{2}X)$ and $g(t) = 0$ outside the interval $(-\tfrac{1}{2}T, \tfrac{1}{2}T)$. Denoting the Fourier transforms of these envelope functions by $F(k_x)$ and $G(\omega)$, respectively, we write

$$f(x) \cos(k_c x) = (4\pi)^{-1} \int_{-\infty}^{\infty} F(k_x) \exp(ik_x x) \left[\exp(ik_c x) + \exp(-ik_c x)\right] dk_x$$

$$= (4\pi)^{-1} \int_{-\infty}^{\infty} F(k_x) \{\exp[i(k_x + k_c)x] + \exp[i(k_x - k_c)x]\} dk_x$$

$$= (4\pi)^{-1} \int_{-\infty}^{\infty} [F(k_x - k_c) + F(k_x + k_c)] \exp(ik_x x) \, dk_x. \quad (E1)$$

$$g(t) \cos(\omega_s t) = (4\pi)^{-1} \int_{-\infty}^{\infty} G(\omega) \exp(-i\omega t) \left[\exp(i\omega_s t) + \exp(-i\omega_s t)\right] d\omega$$

$$= (4\pi)^{-1} \int_{-\infty}^{\infty} G(\omega) \{\exp[-i(\omega - \omega_s)t] + \exp[-i(\omega + \omega_s)t]\} d\omega$$

$$= (4\pi)^{-1} \int_{-\infty}^{\infty} [G(\omega + \omega_s) + G(\omega - \omega_s)] \exp(-i\omega t) \, d\omega. \quad (E2)$$

The product of the above functions of $x$ and $t$ yields a compact wave-packet $p(x, t)$, as follows:

$$p(x, t) = f(x) \cos(k_c x) \, g(t) \cos(\omega_s t)$$

$$= (4\pi)^{-2} \iint_{-\infty}^{\infty} [F(k_x - k_c) + F(k_x + k_c)][G(\omega + \omega_s) + G(\omega - \omega_s)] \exp[i(k_x x - \omega t)] \, dk_x d\omega. \quad (E3)$$

The compact packet $p(x, t)$ is seen to consist of two groups of plane-waves, namely,

i) those that propagate along the positive $x$-axis, for which $k_x$ and $\omega$ are both positive or both negative, i.e., plane-waves residing in the first and third quadrants of the $k_x \omega$-plane;

ii) those that propagate along the negative $x$-axis, where $k_x$ and $\omega$ have opposite signs, i.e., plane-waves residing in the second and fourth quadrants of the $k_x\omega$-plane.

The superposition of each group of plane-waves exclusively by themselves creates a real-valued function of $x$ and $t$, say $p^{(+)}(x,t)$ for an up-going wave, and $p^{(-)}(x,t)$ for a down-going wave along the $x$-axis. There is no guarantee, however, that either $p^{(+)}$ or $p^{(-)}$ will have a finite duration in time and/or a finite width in space. Nevertheless, if we split the term $\cos(k_c x)\cos(\omega_s t)$ on the left-hand side of Eq.(E3) into $\cos(k_c x - \omega_s t)$ and $\cos(k_c x + \omega_s t)$, then retain only the first term, we will obtain a new wave-packet that not only is real-valued but also has finite duration in time and finite width along the $x$-axis, as follows:

$$p(x,t) = f(x)g(t)\cos(k_c x - \omega_s t)$$
$$= \tfrac{1}{8\pi^2}\iint_{-\infty}^{\infty}[F(k_x - k_c)G(\omega - \omega_s) + F(k_x + k_c)G(\omega + \omega_s)]\exp[\mathrm{i}(k_x x - \omega t)]\,\mathrm{d}k_x\mathrm{d}\omega. \quad (E4)$$

The trouble with this modified $p(x,t)$ is that its Fourier transform now contains plane-waves in the 2nd and 4th quadrants of the $k_x\omega$-plane, which propagate along the negative $x$-axis. One should be able to live with these plane-waves, however, so long as $F(k_x)$ and $G(\omega)$ are relatively narrow functions of their respective arguments, and that the shifts (i.e., by $\pm k_c$ along the $k_x$-axis, and by $\pm\omega_s$ along the $\omega$-axis) are large enough to substantially diminish the amplitudes of the plane-waves residing in the 2nd and 4th quadrants.

**References**

1. M. Mansuripur and P. K. Jakobsen, "Theoretical analysis of Fresnel reflection and transmission in the presence of gain media," *Optical Review*, published online: 09 August, 2021; https://doi.org/10.1007/s10043-021-00690-4.
2. M. Born and E. Wolf, *Principles of Optics*, 7th (expanded) edition, Cambridge University Press, Cambridge, United Kingdom (1999).
3. J. D. Jackson, *Classical Electrodynamics*, 3rd edition, Wiley, New York (1999).
4. F. B. Hildebrand, *Advanced Calculus for Applications*, 2nd edition, Prentice-Hall, Upper Saddle River, New Jersey (1976).
5. G. B. Arfken and H. J. Weber, *Mathematical Methods for Physicists*, 6th edition, Academic Press, Cambridge, Massachusetts (2005).
6. M. Mansuripur, *Mathematical Methods in Science and Engineering: Applications in Optics and Photonics*, Cognella Academic Publishing, San Diego, California (2020).
7. A. E. Siegman, *Lasers*, University Science Books, Sausalito, California (1986).
8. M. Mansuripur, *Field, Force, Energy and Momentum in Classical Electrodynamics*, revised edition, Bentham Science Publishers, Sharjah (2017).
9. R. N. Bracewell, *The Fourier Transform and Its Applications*, 2nd edition, McGraw-Hill, New York (1978).
10. R. J. Briggs, *Electron-Stream Interactions with Plasmas*, MIT Press, Cambridge, Massachusetts (1964).
11. J. Skaar, "Fresnel equations and the refractive index of active media," *Physical Review E* **73**, 026605 (2006).
12. B. Nistad and J. Skaar, "Causality and electromagnetic properties of active media," *Physical Review E* **78**, 036603 (2008).
13. H. O. Hågenvik and J. Skaar, "Laplace-Fourier analysis and instabilities of a gainy slab," *Journal of the Optical Society of America B* **32**, 1947-53 (2015).
14. T. S. Mansuripur and M. Mansuripur, "Fresnel reflection from a cavity with net roundtrip gain," *Applied Physics Letters* **104**, 121106 (2014).
15. A. E. Siegman, "Fresnel Reflection, Lenserf Reflection, and Evanescent Gain," *Optics & Photonics News*, pp38-45 (January 2010).